\documentclass{article}
\usepackage[utf8]{inputenc}
\usepackage{graphicx}
\usepackage{color}
\usepackage{multirow}
\usepackage{latexsym}
\usepackage{amsmath,array}
\usepackage{tabularx}
\usepackage{soul}
\usepackage{tabu}

\title{Price-Wharton Constrained Colliders: Co-Causation or No Causation?}

\author{W.M. Stuckey\thanks{Department of Physics, Elizabethtown College, Elizabethtown, PA 17022, USA} \, and Michael Silberstein\thanks{Department of Philosophy, Elizabethtown College, Elizabethtown, PA 17022, USA} \, \thanks{Department of Philosophy, University of Maryland, College Park, MD 20742, USA}}

\date{}

\begin{document}

\maketitle
\sloppy
\begin{abstract}
\noindent Price and Wharton have recently suggested that ``constrained retrocausal collider bias is the origin
of entanglement.'' In this paper, we argue that their \textit{connection across a constrained collider} (CCC) for the V-shaped case with the Bell states is not ``a mechanism for entanglement,'' providing a negative answer to the title of https://arxiv.org/abs/2406.04571. Rather, CCC should be viewed as a novel approach to the causal modeling of entanglement, providing a perspectival co-causal relationship between the two wings of the experiment that does not violate locality or statistical independence. The key to this advance in the causal modeling of entanglement is to accept that quantum mechanics is complete, so the Bell states provide the ``mechanism for entanglement'' and CCC provides a causal model of entanglement per causal perspectivalism in accord with the different subjective spacetime models of the experiment. When combined into an objective spacetime model of the experiment, the subjective co-causation disappears leaving an objective acausal view of entanglement in accord with the axiomatic reconstruction of quantum mechanics via information-theoretic principles (quantum reconstruction program). Essentially, the quantum reconstruction program has rendered quantum mechanics a principle theory based on the observer-independence of Planck's constant $h$ as justified by the relativity principle (``no preferred reference frame'' NPRF), exactly as special relativity is a principle theory based on the observer-independence of the speed of light $c$ as justified by the relativity principle. Thus, NPRF + $c$ is an \textit{adynamical global constraint} on the configuration of worldtubes for the equipment in the experiment while NPRF + $h$ is an adynamical global constraint on the distribution of quantum detection events in that context. Accordingly, CCC then provides a new way to understand causation subjectively per causal perspectivalism in this objectively acausal spacetime model of the experiment.
\end{abstract}

\clearpage


\section{Introduction}\label{SectionIntro}

In recent papers \cite{PriceColliders2024}\cite{pricewhartonFlipV2015}\cite{PriceWhartonColliders2022}\cite{PriceWhartonColliders2024}, Price and Wharton inverted and constrained a collider from causal modeling theory to explain Bell-inequality-violating correlations in co-(retro)causal fashion (Section \ref{SectionZigZag}). According to Price \cite{PriceColliders2024}, ``This hypothesis requires no direct causal influence outside lightcones, and may hence offer a new way to reconcile Bell nonlocality and relativity.'' Since a Bell state distribution of quantum events in spacetime is independent of the spatial separation and temporal order of the two measurement events in each trial of the experiment, the joint probabilities (and hence the correlations) are invariant under Lorentz boosts. Thus, unless one invokes an additional (hidden) causal mechanism in an attempt to account for Bell-inequality-violating correlations constructively, there is no reason to suspect a conflict with special relativity (SR). 

Indeed, Bell's theorem proves that any constructive account of Bell-inequality-violating correlations must violate locality, statistical independence, intersubjective agreement, and/or unique experimental outcomes \cite{EinsteinsEntanglement2024}. Nonetheless, the overwhelming majority of physicists and philosophers working in the foundations of quantum mechanics (foundations community) posit and/or champion constructive accounts of Bell state entanglement in accord with Reichenbach's Principle \cite{ReichenbachSEP}:
\begin{quote}
If A and B are correlated, then A is a cause of B; B is a cause of A; or A and B are both caused by a third factor, C. In the last case, the common cause C occurs prior to A and B.
\end{quote}
Given the dearth of alternative structural or principle accounts, this is not entirely unreasonable. And, given that our subjective experience is dynamical, it is natural to seek constructive explanations in general. As Allori wrote \cite{allori2019}:
\begin{quote}
 Philosophers of science have proposed objective accounts of explanation, but they all recognize there's a strong sense in which explanation is ``explanation for us,'' and any account should capture our intuition that explanation is fundamentally dynamical. This is connected with causation: intuitively, we explain an event because we find its causes; causes happen before their effects and ``bring them about.''
\end{quote} 
Of course, retrocausation violates the expectation that ``causes happen before their effects,'' but that is perhaps an archaic notion of causation associated with Newtonian thinking \cite{adlam2021}. We have presented a longer defense of thinking about causation beyond simple Newtonian mechanics elsewhere \cite{EinsteinsEntanglement2024}, so here we will proceed straight to an analysis of Price-Wharton co-causal colliders. 

\begin{figure}
\begin{center}
\includegraphics [height = 60mm]{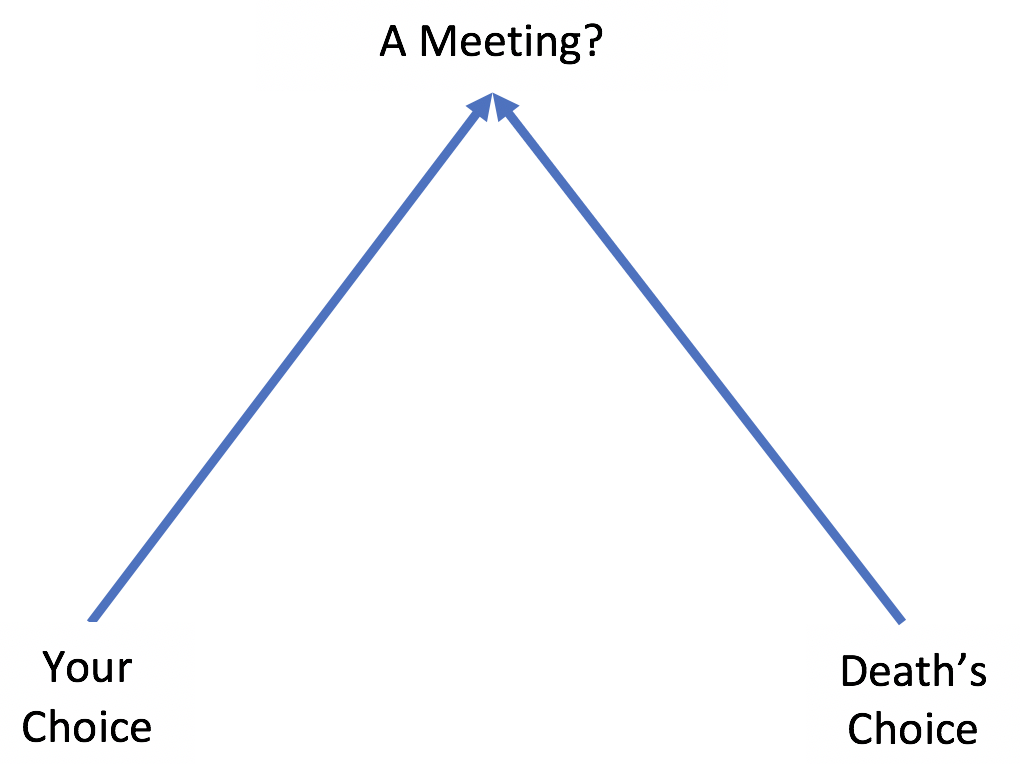}  \caption{Death in Damascus collider \cite{PriceWhartonColliders2024}.} \label{DeathDamascus}
\end{center}
\end{figure}

A \textit{collider} is simply a node in a causal modeling graph where two (or more) causes meet. For example, in \cite{PriceWhartonColliders2024} Price and Wharton illustrate a simple collider with the ``Death in Damascus case'' (Figure \ref{DeathDamascus}). Suppose you and Death have two options for where to travel tomorrow, Damascus or Aleppo, and if you choose to go to the same city that Death chooses (collider value 1), you die. If you choose different cities (collider value 0), you live. If many people participate in this activity and I talk to all those I can find the next day, I discover that they all possess an uncanny knack for avoiding Death. But this is just a selection bias, since I can't interview those who died. Indeed, some of those I interview might believe they really do possess a special quality of `survivorness', so that had they chosen differently, Death would also have chosen differently. 

It is normal to reject this belief, i.e., we tend to believe that had a survivor chosen the other city, they would have met Death and died. But, suppose Fate intervenes and constrains the collider value to be 0 in your case. If you are free to choose whichever city you like, then Death is not. In this sense, a constrained collider allows you to control Death's movement. Of course, Death might argue equally that it is controlling your movement. So, we see that a constrained collider allows for a notion of \textit{co-causality}. Of course, as with the `survivorness' quality, it is normal to reject the intervention of Fate, but this will become relevant for the inverted collider of interest.

\begin{figure}
\begin{center}
\includegraphics [height = 50mm]{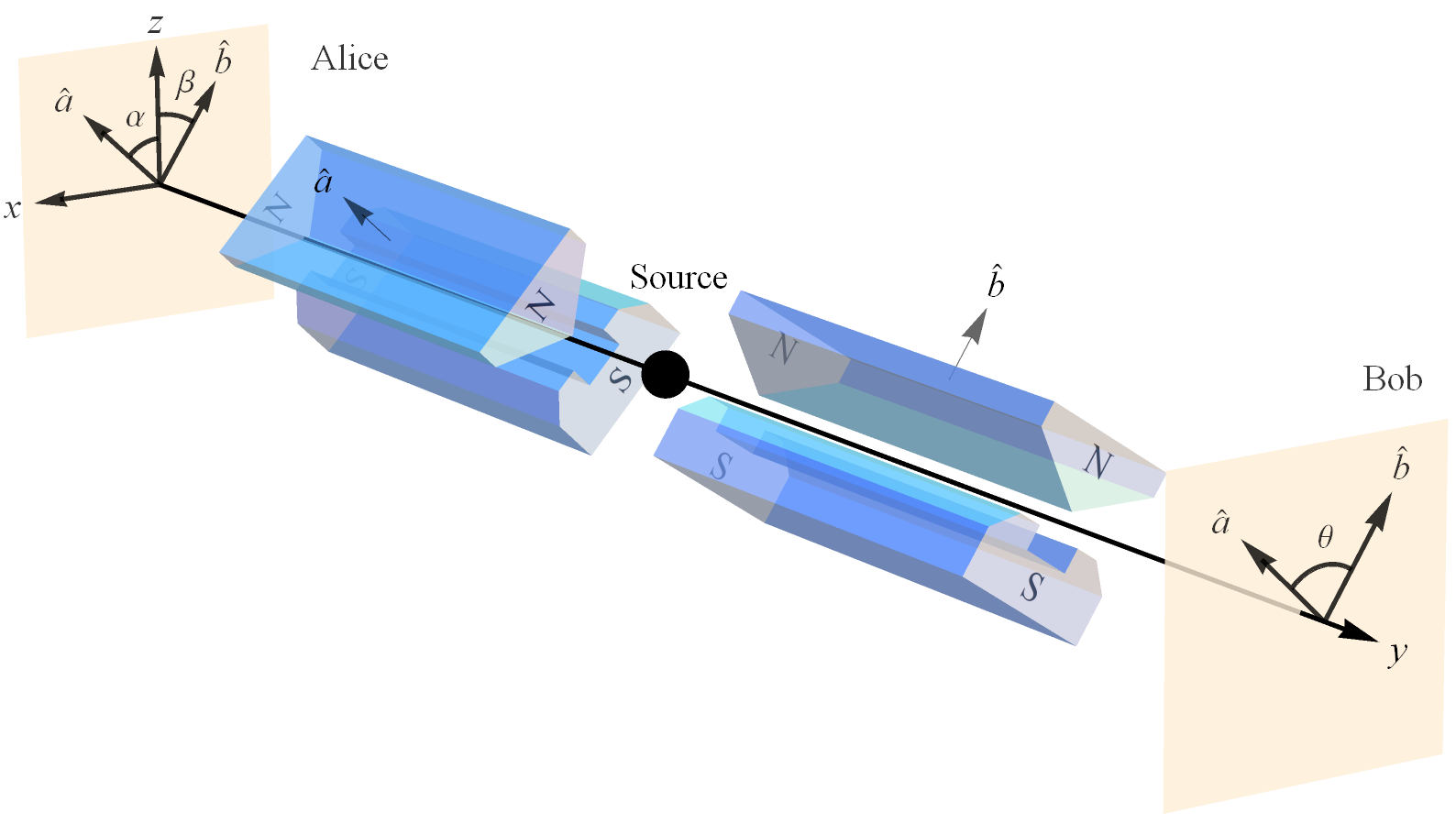}  \caption{Alice and Bob making spin measurements on a pair of spin-entangled particles with their Stern-Gerlach (SG) magnets and detectors.} \label{EPRBmeasure}
\end{center}
\end{figure}

\begin{figure}
\begin{center}
\includegraphics [width=\textwidth]{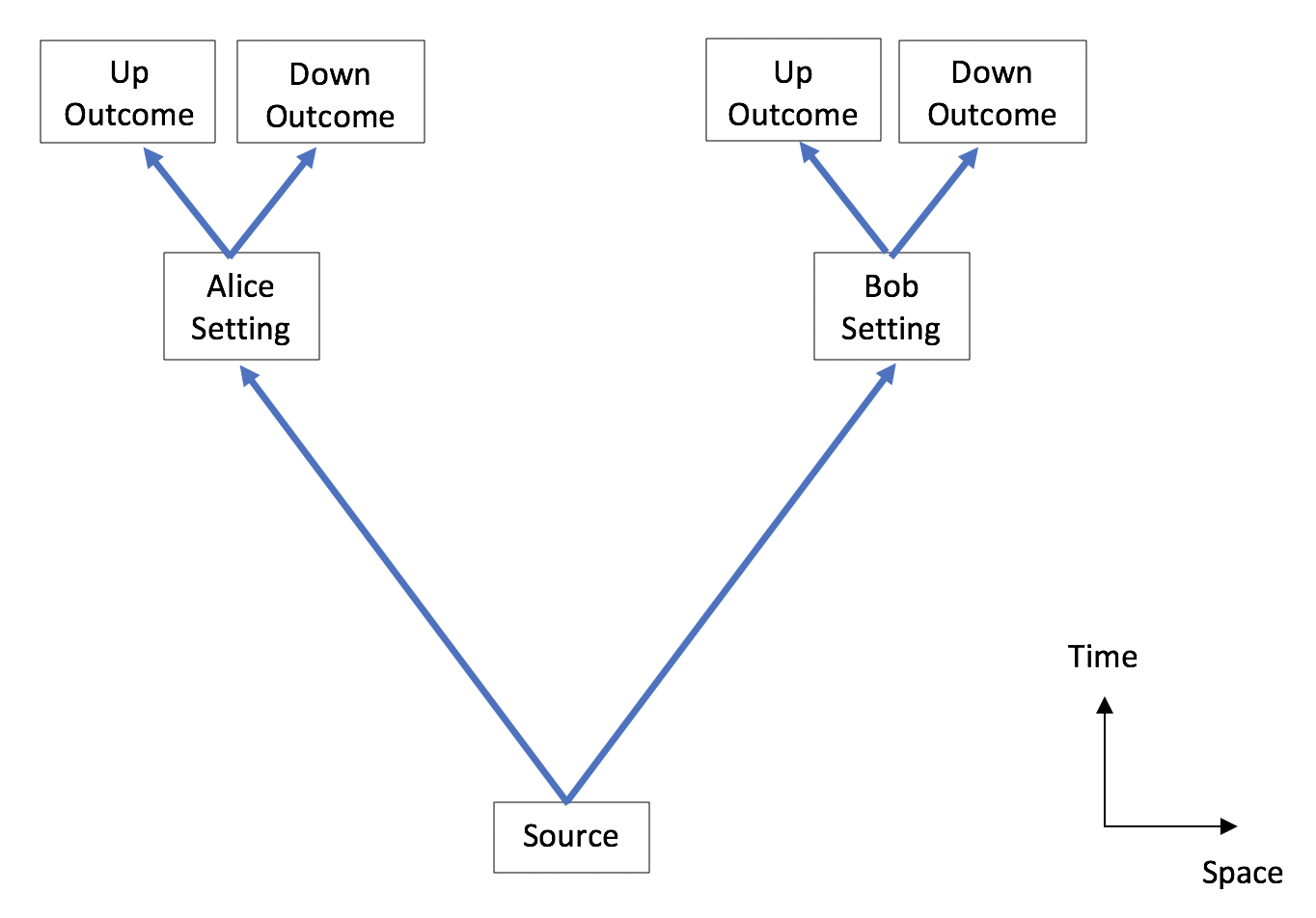}  \caption{The V-Shaped Case. The blue arrows depict two particles leaving the Source towards Alice and Bob's SG measurement devices. Each trial of the experiment produces one of the two spin outcomes shown above each of their settings.} \label{Vshape}
\end{center}
\end{figure}

\begin{figure}
\begin{center}
\includegraphics [height = 50mm]{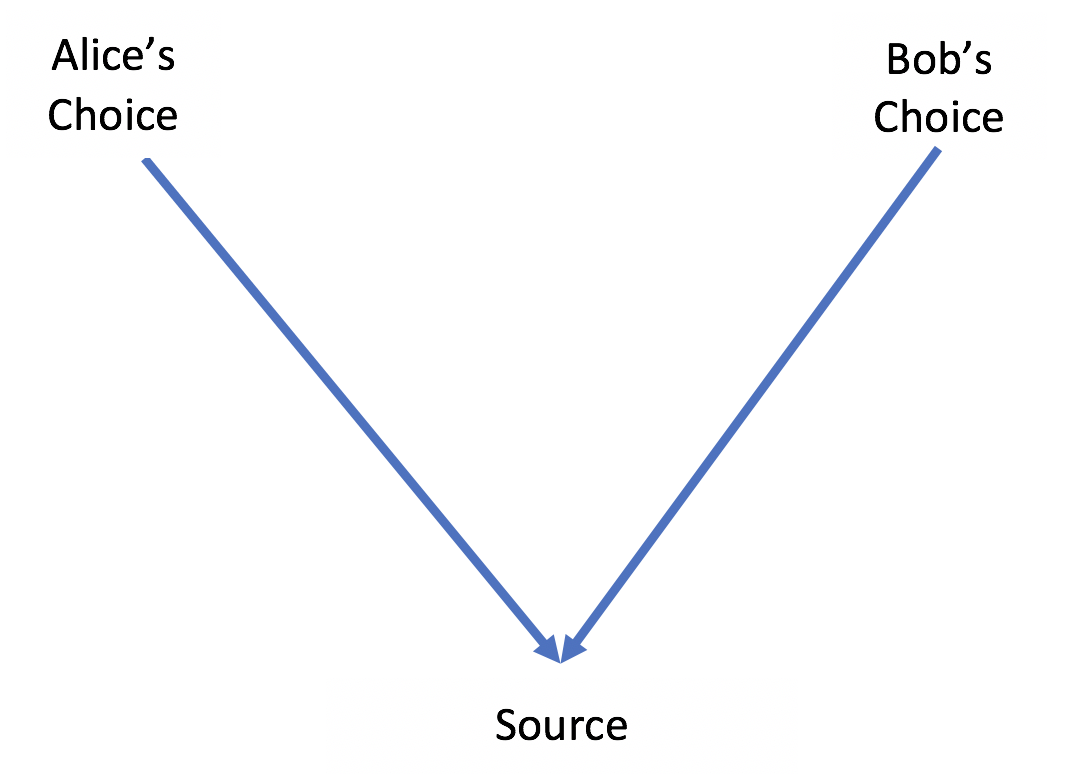}  \caption{Inverted Collider for Bell Spin State Measurements. When the Source is unconstrained, the simple inverted collider graph depicts (retro)causal influences on the Source from Alice and Bob.} \label{BellCollider}
\end{center}
\end{figure}

\begin{figure}
\begin{center}
\includegraphics [width=\textwidth]{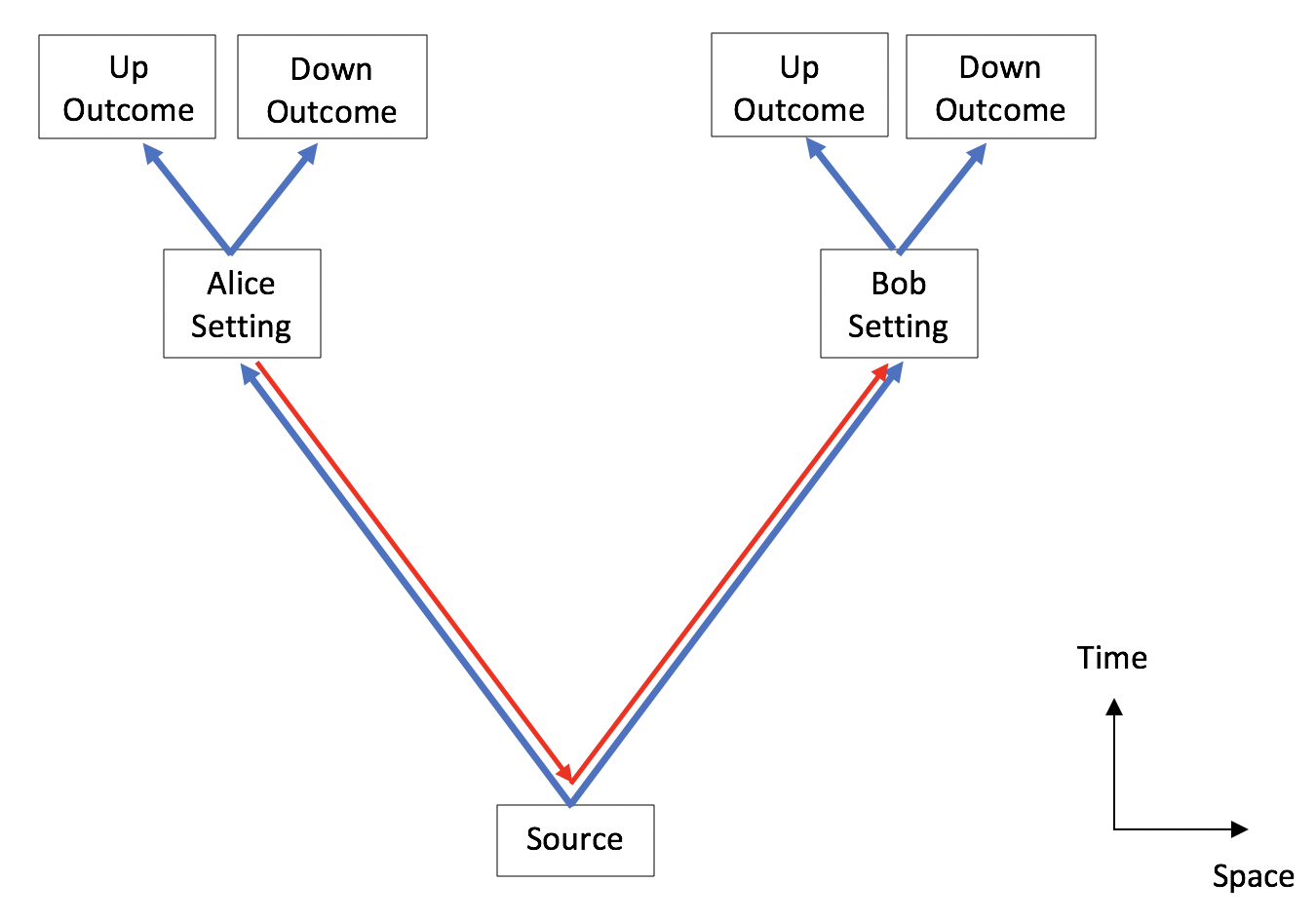}  \caption{Zigzag Pattern in the V-Shaped Case. The red arrows depict a timelike causal ``zigzag'' pattern from Alice's SG measurement setting to the Source then to Bob's SG measurement outcome.} \label{ZigZag1A}
\end{center}
\end{figure}

\begin{figure}
\begin{center}
\includegraphics [width=\textwidth]{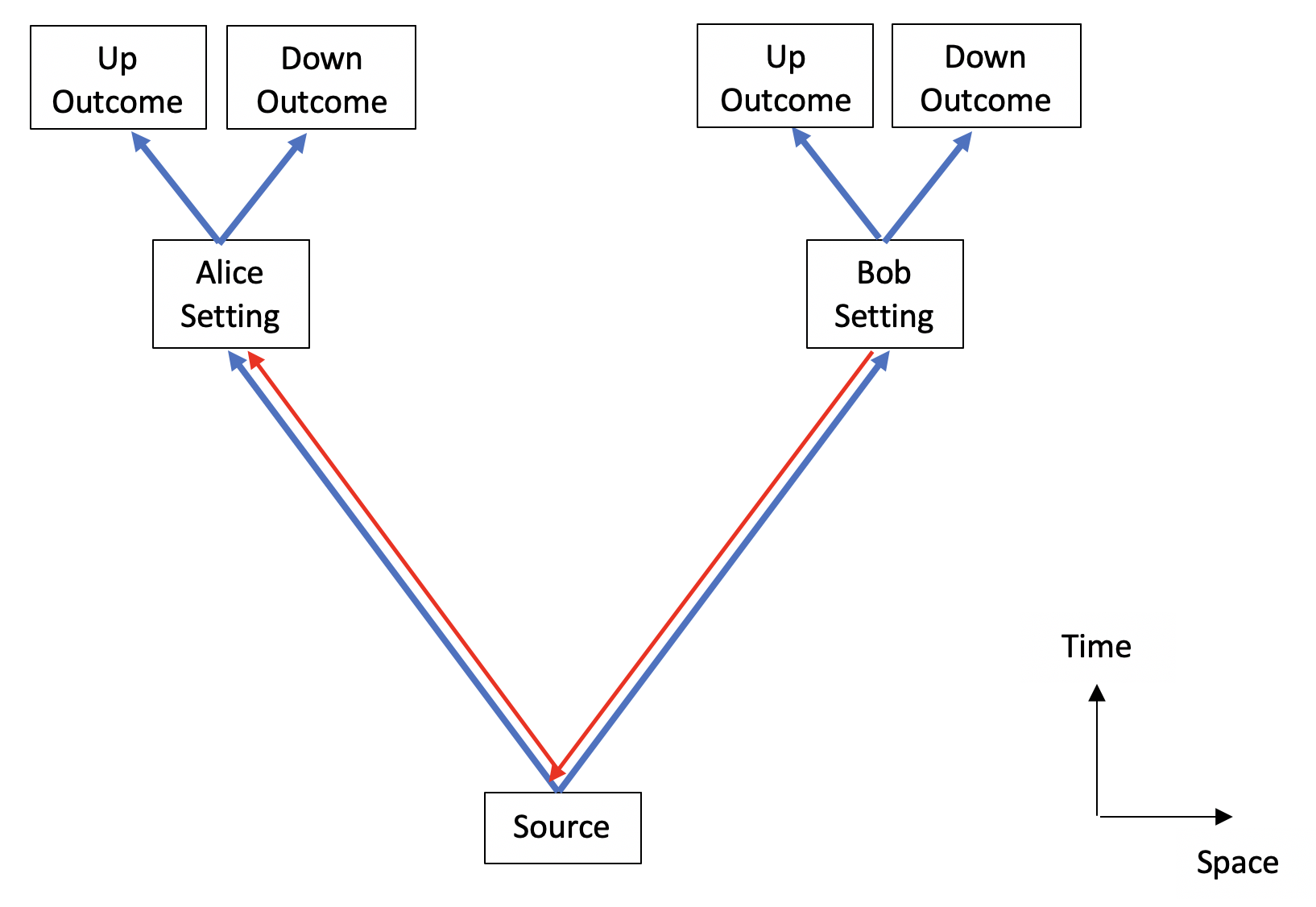}  \caption{Zigzag Pattern in the V-Shaped Case. The red arrows depict a timelike causal ``zigzag'' pattern from Bob's SG measurement setting to the Source then to Alice's SG measurement outcome.} \label{ZigZag1}
\end{center}
\end{figure}

That is, let's invert the simple collider and apply it to the V-shaped case for Bell spin state measurements (Figures \ref{EPRBmeasure} and \ref{Vshape}). Suppose the source is unconstrained, as was the case for our Death in Damascus collider without Fate. In that case, the inverted collider graph gives us retrocausation from Alice and Bob to the Source (Figure \ref{BellCollider}). Of course, the source is constrained in such experiments, i.e., the experimentalist goes to great lengths to create a particular Bell state source, e.g., see \cite{dehlinger}. But, that means the applicable inverted collider model is for a constrained collider a la our addition of Fate for Death in Damascus. And that means Alice (Figure \ref{ZigZag1A}) and Bob (Figure \ref{ZigZag1}) can both claim causal influence on the opposite wing of the experiment, just like you and Death in the Fate constrained Death in Damascus collider. Thus, this \textit{connection across a constrained collider} (CCC) shows how ``measurement choices on one side of an Bell experiment \textit{make a difference} to outcomes on the other side, in some cases'' \cite{PriceWhartonColliders2024}. 

This is the sense in which Price and Wharton propose CCC as a ``mechanism for entanglement.'' Of course, everything said about the co-causal collider would apply equally to a source of classical particles with counterfactually definite properties. In that case, the explanation of the non-Bell-inequality-violating correlations would follow the blue arrows in Figure \ref{Vshape} (for non-spin, classical particles) trivially. That is, there would be no reason to invoke the co-causality of CCC with its hidden co-causal mechanism violating statistical independence for such a source. Therefore, we answer the title ``A Mechanism for Entanglement?'' of \cite{PriceWhartonColliders2024} in the negative. Instead, we will argue for an entirely different role for CCC in Bell state entanglement.

To make that argument, we will compare and contrast Price and Wharton's causal (constructive) account of Bell state entanglement via CCC with our acausal (principle) account via the axiomatic reconstruction of quantum mechanics (QM). Specifically, the quantum reconstruction program has successfully rendered QM a \textit{principle theory} \cite{GoyalPhenomQBism2024} in Einstein's sense of the term (Section \ref{SectionPrinciple-alternative}). According to Einstein, the formalism of a principle theory is derived from an empirically discovered fact, e.g., the laws of thermodynamics from the empirically discovered fact that ``perpetual motion machines are impossible'' \cite{einstein1919}. The empirically discovered fact that the quantum reconstruction program used to derive the (finite-dimensional) Hilbert space (kinematic structure) of QM is Information Invariance \& Continuity.  

\clearpage

Since Information Invariance \& Continuity is an information-theoretic principle, its physical meaning is not as transparent as, say, the empirically discovered fact whence the Lorentz transformations of SR, i.e., the light postulate. The light postulate states that everyone measures the same value for the speed of light $c$, regardless of their uniform relative motions (the observer-independence of $c$). In order to render Information Invariance \& Continuity at the basis of the principle theory of QM as physically transparent as the light postulate at the basis of the principle theory of SR, we spatialized the otherwise abstract notion of measurement in the quantum reconstruction program \cite{NPRF2022}\cite{EinsteinsEntanglement2024}. As a result, Information Invariance \& Continuity is seen to entail the observer-independence of Planck's constant $h$. That is, everyone measures the same value for Planck's constant $h$, regardless of their relative spatial orientations or locations (Planck postulate). 

In SR, the justification of its empirically discovered fact (light postulate) is the relativity principle -- the laws of physics (including their constants of Nature) are the same in all inertial reference frames. That's because the value $c$ in the light postulate is part of Maxwell's equations. We abbreviate this NPRF + $c$, where NPRF stands for ``no preferred reference frame.'' Obviously, we can also use the relativity principle to justify the Planck postulate, since spatial rotations and translations also relate inertial reference frames and Planck's constant $h$ is a constant of Nature per Planck's radiation law. In analogy with SR, we abbreviate this NPRF + $h$. 

After comparing co-causality per CCC with acausality per NPRF + $h$ for Bell state entanglement, we make the following analogy with SR (Sections \ref{SectionSR} and \ref{SectionCo-causation}). Suppose you believe that CCC requires some unknown completion of QM \cite{WhartonEmail2024} in analogy with the luminiferous aether for SR. Accordingly, length contraction is the result of some (hidden) causal mechanism whereby Alice (Bob) can claim Bob's (Alice's) meter sticks are literally shorter than hers (his). Instead of attributing length contraction to a dynamic effect, physicists by and large rather believe it to be a kinematic fact resulting from a principle constraint (NPRF + $c$). 

Likewise per CCC, the `average-only' conservation of Bell state entanglement is the result of some (hidden) co-causal mechanism violating statistical independence, whereby Alice (Bob) can claim Bob's (Alice's) data must be averaged to satisfy Bell state conservation. Since we now know that `average-only' conservation can be understood as a kinematic fact resulting from a principle constraint (NPRF + $h$), this analogy strongly suggests that the zigzag co-causation of CCC is just as superfluous for QM as the luminiferous aether is for SR. 

Thus, in Section \ref{SectionConcl} we propose an alternative view of CCC for Bell state entanglement. First, we suggest the advocate of CCC simply accept the completeness of QM, so that the analogy with the constructive completion of SR fails. Second, CCC is viewed as the subjective, causal counterpart to the objective, acausal NPRF + $h$ per causal perspectivalism \cite{evans2015}\cite{price2007}\cite{price2010}. That is, the red causal zigzags in Figures \ref{ZigZag1A} and \ref{ZigZag1} are certainly valid for Alice and Bob's subjective spacetime models (causal perspectivalism) and disappear when combining the co-causal subjective spacetime models into an acausal objective spacetime model. This acknowledges the dynamical nature of subjective experience responsible for the subjective spacetime models of reality, which are sine qua non for producing an objective spacetime model. Indeed, the subjective spacetime models are co-fundamental with the objective spacetime model in this view of physics \cite{EinsteinsEntanglement2024}. Accordingly, far from being superfluous, CCC is a novel causal account of Bell state entanglement that does not violate locality or statistical independence. In conclusion, NPRF + $c$, NPRF + $h$, and CCC provide a complete explanans for Bell state entanglement that does not violate locality, statistical independence, intersubjective agreement, or the uniqueness of experimental outcomes, and they reveal QM to be as complete as possible.


\section{Can We Save Reichenbach?}\label{SectionZigZag}

\begin{figure}
\begin{center}
\includegraphics [height = 70mm]{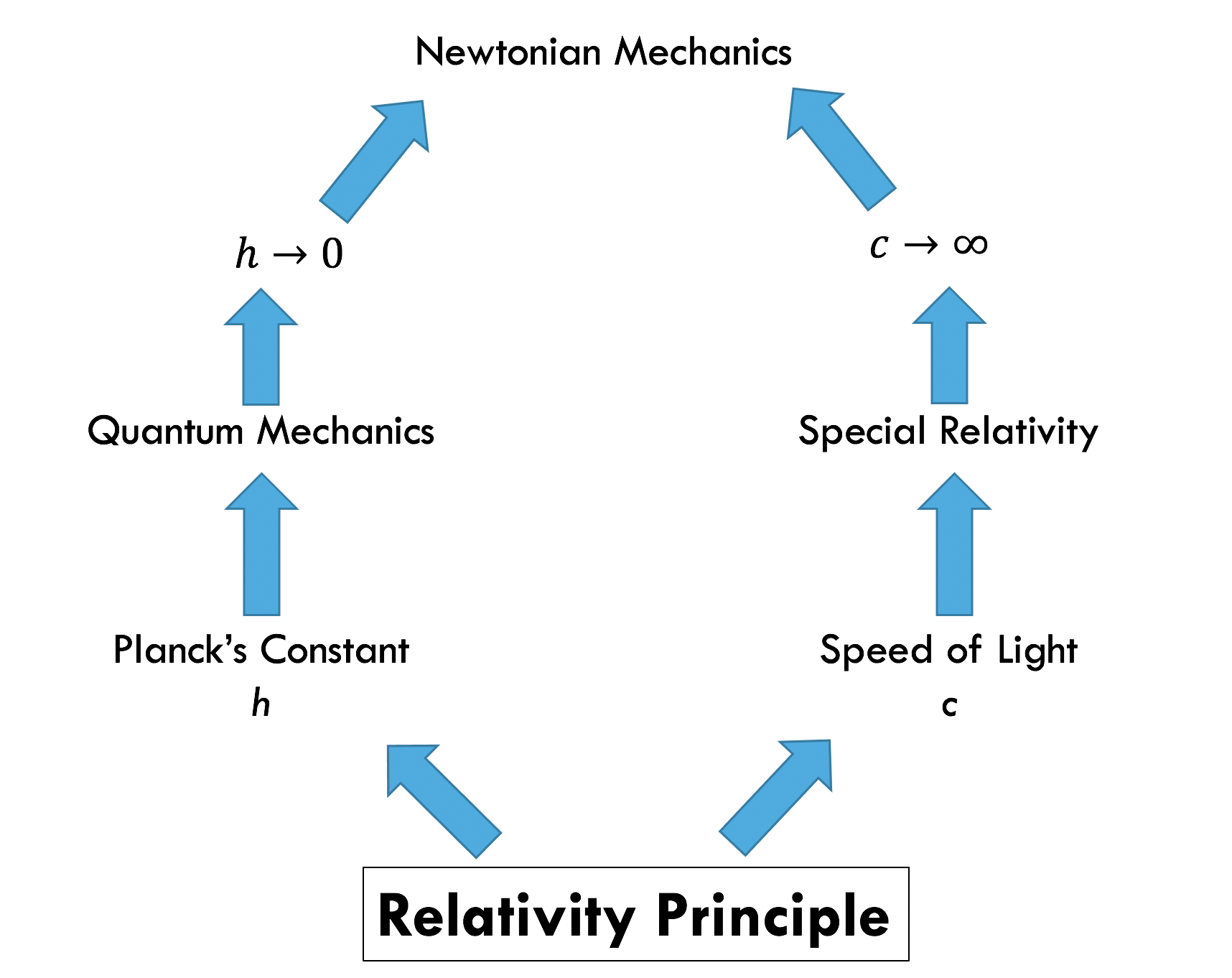}  \caption{The variables in Newtonian mechanics commute which means $h \rightarrow 0$ in the commutator for the corresponding variables in quantum mechanics. Newtonian equations hold on average according to QM and follow from the corresponding equations in SR with $c \rightarrow \infty$.} \label{QM-SR-Newton}
\end{center}
\end{figure}

Clearly, Bell's theorem and the experimental violation of Bell's inequality tell us Reichenbach's Principle fails for the correlations of quantum entanglement, at least if one wants causality along the lines of Newtonian mechanics in accord with locality, statistical independence, intersubjective agreement, and unique experimental outcomes \cite{EinsteinsEntanglement2024}. But, restricting the notion of causality to that of Newtonian mechanics is a rather simplistic view of causality. Plus, we know that Newtonian mechanics is not a fundamental theory of physics (Figure \ref{QM-SR-Newton}). As Adlam stated \cite{adlam2021}:
\begin{quote}
there have been several major conceptual revolutions in physics since the time of Newton\index{Newton, Isaac}, and thus we should not necessarily expect that accounts of lawhood based on a Newtonian time-evolution picture will be well-suited to the realities of modern physics.
\end{quote}
Perhaps there are more sophisticated ways to understand causality in accord with the more fundamental theory of QM, so as to salvage Reichenbach's Principle in some sense and perhaps legitimize the dynamical nature of subjective experience. 

One can certainly provide accounts of causation that are consistent with the weirdness of the quantum, such as interventionist or manipulability accounts of causation \cite{pearl2009}. The central idea is that X is a cause of Y if and only if manipulating X is an effective means of indirectly manipulating Y. According to retrocausal accounts of QM espousing an interventionist account of causation, manipulating the setting of a measurement apparatus now can be an effective means of manipulating aspects of the past. The formal machinery of causal modelling has the interventionist account of causality as its foundation. For the V-shaped case with Alice and Bob making Stern-Gerlach (SG) measurements on a Bell spin state (Figures \ref{EPRBmeasure} and \ref{Vshape}), that means the Bell state correlations depend on Bob's SG measurement settings, so Bob has intervened in these correlations. Likewise, the Bell state correlations depend on Alice's SG measurement settings, so Alice has intervened in these correlations.  

\clearpage

Price and Wharton, two key defenders of retrocausal accounts of the quantum, embrace a subset of interventionism known as the agent or perspectivalism account of causation \cite{evans2015}\cite{price2007}\cite{price2010}. On this view, causal relations are relations that can be used for control or manipulation, from the perspective of the agent in question of course. This is an understandably appealing notion of causation for those such as Price and Wharton who espouse a block universe picture, wherein causation talk cannot possibly be about changing or bringing about events (past, present or future) in any robust sense of those terms. So, such accounts of causation do not provide dynamical explanations for Bell-inequality-violating correlations in any objective sense, but we can understand them as attempts to save the appearances of causal explanations in accord with our dynamical subjective experience.

We should point out that retrocausality does not allow one to send messages (signals) into the past, so Wharton and Argaman \cite{whartonArgaman2020} use the term ``future-input dependence'' rather than ``retrocausality.'' Argaman writes \cite{argaman2018}: ``Signaling into the past must thus indeed be strictly impossible, as it would allow construction of causal loops – the well-known inconsistency arguments of the grandfather paradox (a.k.a. the bilking argument).'' Similarly, Wharton and Liu \cite{WhartonLiu2022} used the path integral formalism for QM that ``is much more evidently time-symmetric than conventional [quantum mechanics], bringing it closer to a causally-neutral account.'' 

Adlam calls retrocausality that is viewed in this block universe fashion \textit{all-at-once retrocausality} and contrasts that with \textit{dynamical retrocausation} which is ``mediated by physical systems travelling backwards in time'' \cite{adlam2022}. Specific instantiations of dynamical retrocausality include Aharonov, Bergmann and Lebowitz's two-states vector formalism \cite{aharonov1} and Cramer's transactional interpretation \cite{cramer}. However, Cramer admits the backward-causal elements of his transactional interpretation are ``only a pedagogical convention,'' and that in fact ``the process is atemporal'' \cite{cramer2015}. 

For the V-shaped case, all-at-once retrocausality is often described as making a timelike ``zigzag'' pattern in spacetime between the two spacelike separated measurement events \cite{pricewhartonFlipV2015}\cite{PriceWhartonColliders2022}, as shown in Figures \ref{ZigZag1A} and \ref{ZigZag1}. The main complaint about timelike zigzag causation is that events are not being brought into existence in accord with our everyday intuition because part of the zigzag causal influence is coming from future events that have never existed. But, the belief that those future events have never existed is in accord with the Newtonian spacetime model of reality, which is challenged on the coexistence of bodily objects and events by the Minkowskian spacetime model of reality for SR (Section \ref{SectionPrinciple-alternative}). So, let's consider a notion of causality that isn't necessarily Newtonian in that sense. 

In Figure \ref{ZigZag1} for example, Bob's SG setting creates a causal influence backward in time to the Source then forward in time to causally influence the particle-detector interaction at Alice's SG setting. If Alice and Bob's setting events were timelike related, one could simply invoke a forward-time-directed causal influence from the earlier event directly to the later event. Since Alice and Bob's setting events are spacelike related, a zigzag causal influence is needed to avoid violating locality. And, the opposing temporal directions of that causal influence mean it is all-at-once aka atemporal; no events are \textit{objectively} bringing other events into existence. This is consistent with the fact that we could have drawn the zigzag from Alice's SG setting to Bob's and none of the empirical facts would have `changed' in our all-at-once zigzag diagram (Figure \ref{ZigZag1A}). Price and Wharton have shown that this view of causal perspectivalism has a basis in causal modeling, i.e., CCC as explained in Section \ref{SectionIntro}. 

\begin{figure}
\begin{center}
\includegraphics [width=\textwidth]{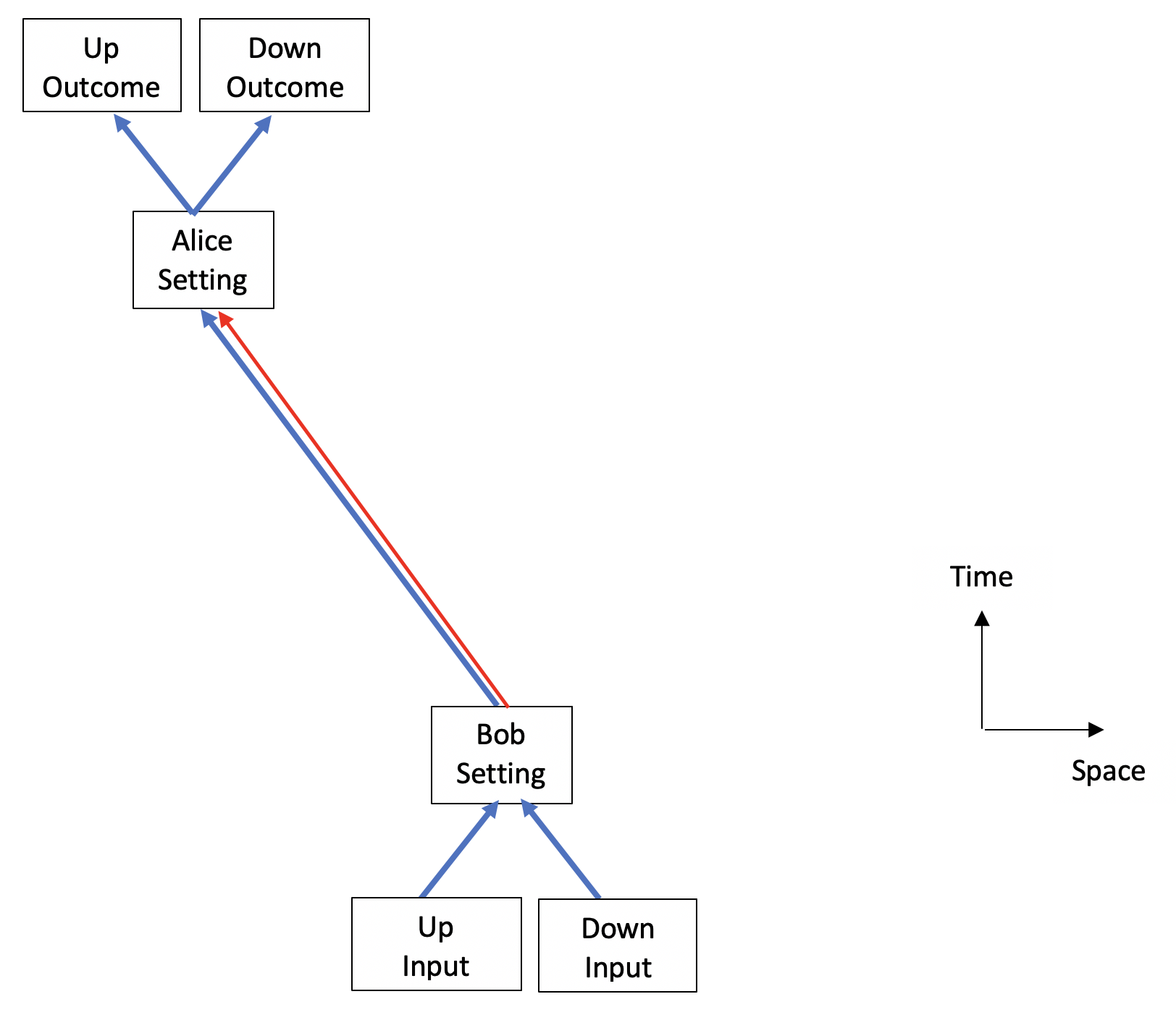}  \caption{We temporally flip Bob's side of a trial for the V-shaped case from that in Figure \ref{ZigZag1}. Now there is only one particle moving forward in time from either of two inputs to Bob's SG magnets then to Alice's SG magnets where we have one of two possible outcomes.} \label{ZigZag2}
\end{center}
\end{figure}

The symmetry of this co-causation distinguishes forward timelike causation from retrocausation and is directly related to quantum phenomena. To see why that is true, consider a different experiment altogether whereby a single particle is input to Bob's SG magnets in one of two states corresponding to up or down for his particular SG magnet setting \cite{pricewhartonFlipV2015}. Now we're sending a particle to Bob's SG magnets with spin up (or down) relative to Bob's SG magnet setting, so of course the particle merely proceeds through his SG magnets with its state unchanged to Alice's SG magnets (Figure \ref{ZigZag2}). We can think of this single-particle experiment as resulting from temporally flipping Bob's half of the experimental trial shown in Figure \ref{ZigZag1} for two particles to produce produce Figure \ref{ZigZag2} for one particle. 

For example, suppose we send a spin up particle to Bob's SG magnets so that it proceeds as a spin up particle with respect to Bob's SG magnets to Alice's SG magnets. There the particle emerges either up ($+1$) or down ($-1$) (in units of $\frac{\hbar}{2}$) with respect to Alice's SG magnets exactly as given by the quantum-mechanical probabilities, i.e., $P(+|\theta) = \cos^2{\left(\frac{\theta}{2}\right)}$ and $P(-|\theta) = \sin^2{\left(\frac{\theta}{2}\right)}$ where $\theta$ is the angle between Bob's SG magnets and Alice's (Figure \ref{EPRBmeasure}). For the situation when we start with a spin down particle, we just have to exchange sine and cosine in the probability functions. In this purely forward timelike sequence of events, we have no trouble saying it is Bob who establishes the cause, i.e., the initial quantum state for Alice to measure. Certainly, there is nothing mysterious here (other than the mystery of quantum superposition per spin itself, see below).

\begin{figure}
\begin{center}
\includegraphics [width=\textwidth]{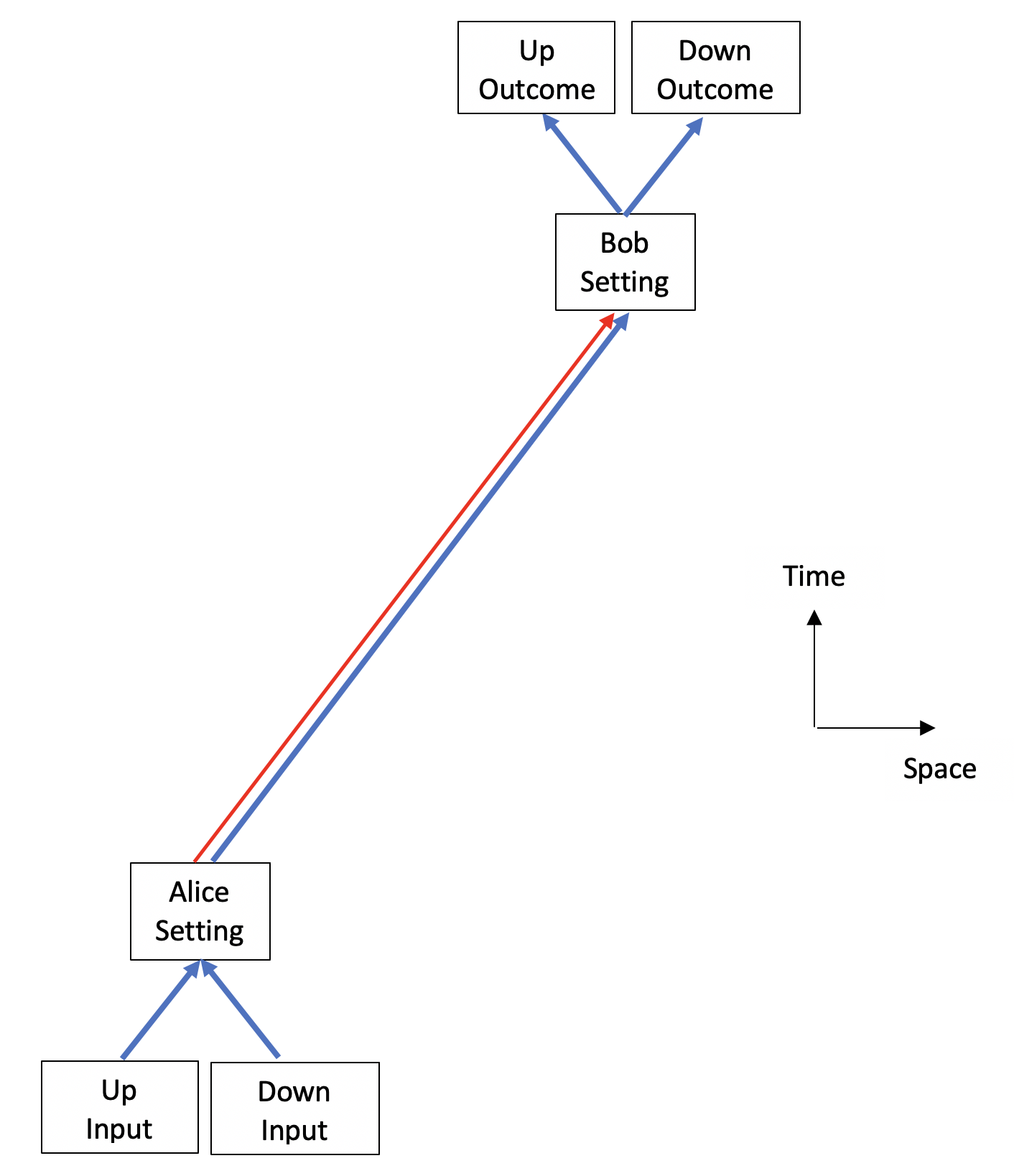}  \caption{We temporally flip Alice's side of a trial for the V-shaped case from that in Figure \ref{ZigZag1A}. Now there is only one particle moving forward in time from either of two inputs to Alice's SG magnets then to Bob's SG magnets where we have one of two possible outcomes.} \label{ZigZag3}
\end{center}
\end{figure}

Next, because the zigzag pattern can go the other way, consider flipping Alice's half of the experimental trial shown in Figure \ref{ZigZag1A} for two particles to produce Figure \ref{ZigZag3} for one particle. Everything we said about Bob can now be said about Alice; that includes the fact that in the temporally flipped case it is clearly Alice who establishes the cause, i.e., the initial quantum state for Bob to measure. So, how is this symmetry due to the quantum nature of spin?

\begin{figure}
\begin{center}
\includegraphics [height = 50mm]{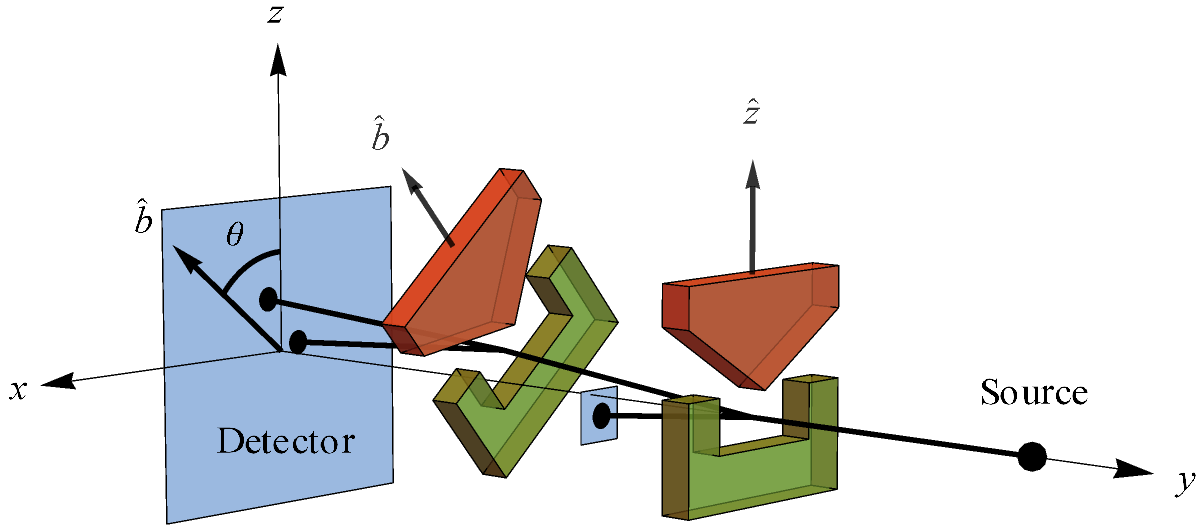}  
\caption{In this set up, the first SG magnets (oriented at $\hat{z}$) are being used to produce an initial spin state $|\psi\rangle = |z+\rangle$ for measurement by the second SG magnets (oriented at $\hat{b}$).} \label{SGExp2}
\end{center}
\end{figure}

\begin{figure}
\begin{center}
\includegraphics [height = 65mm]{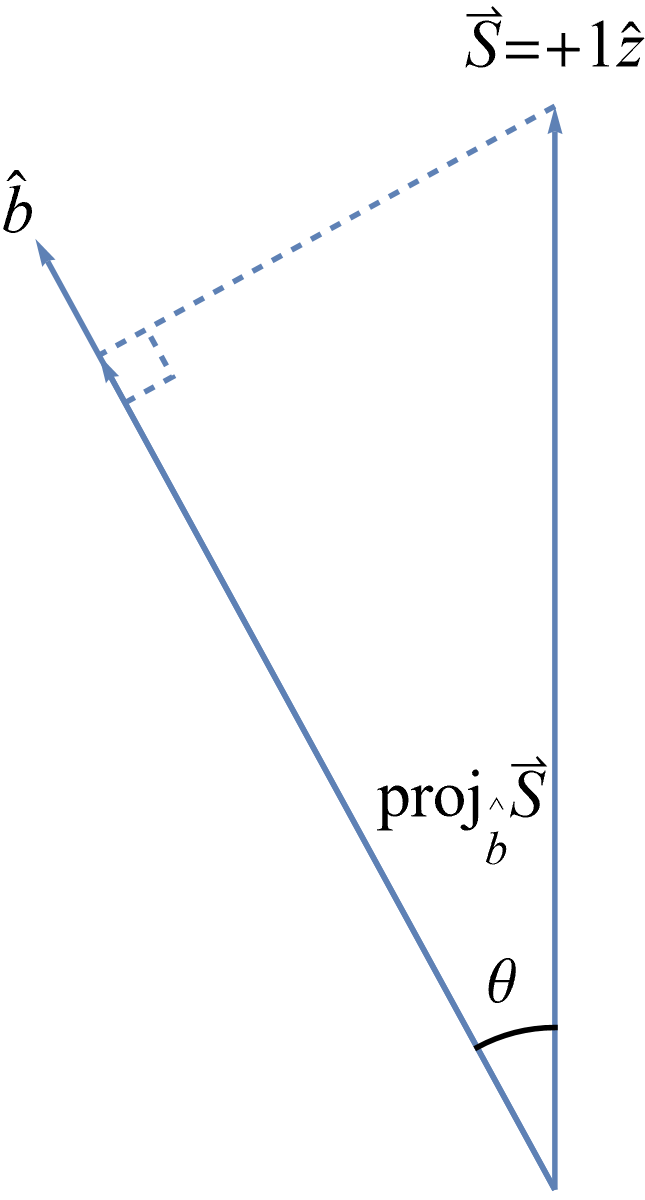} 
\caption{The spin angular momentum $\vec{S}$ projected along the measurement direction $\hat{b}$. This is not what we measure.} \label{Projection}
\end{center}
\end{figure}

Suppose we prepare an initial spin state $|\psi\rangle = |z+\rangle$ for measurement by a subsequent set of SG magnets oriented at $\hat{b}$ (Figure \ref{SGExp2}). Thinking of spin angular momentum as a vector in real space we would expect our SG spin measurement of $|z+\rangle$ along $\hat{b}$ to produce the projection of $+1\hat{z}$ along $\hat{b}$, i.e., $\cos{\left(\theta \right)}$, as shown in Figure \ref{Projection}. But we never observe anything other the two outcomes ``up'' or ``down'' relative to the North magnetic pole, that's the mystery of quantum superposition per the spin-$\frac{1}{2}$ qubit.  

\clearpage

Since, as Weinberg pointed out \cite{weinberg2017}, we are measuring Planck's constant $h$ when we measure the spin of an electron, the mystery of spin amounts to the observer-independence of $h$ between reference frames related by spatial rotations. The quantum reconstruction program has used the information-theoretic generalization of this empirically discovered fact (Information Invariance \& Continuity \cite{brukner2009}) to derive the (finite-dimensional) Hilbert space formalism of QM in total analogy with Einstein's derivation of the Lorentz transformations of SR from the empirically discovered light postulate. As we pointed out in Section \ref{SectionIntro}, both of these empirically discovered facts can be justified by the relativity principle (NPRF). 

\begin{figure}
\begin{center}
\includegraphics [height = 75mm]{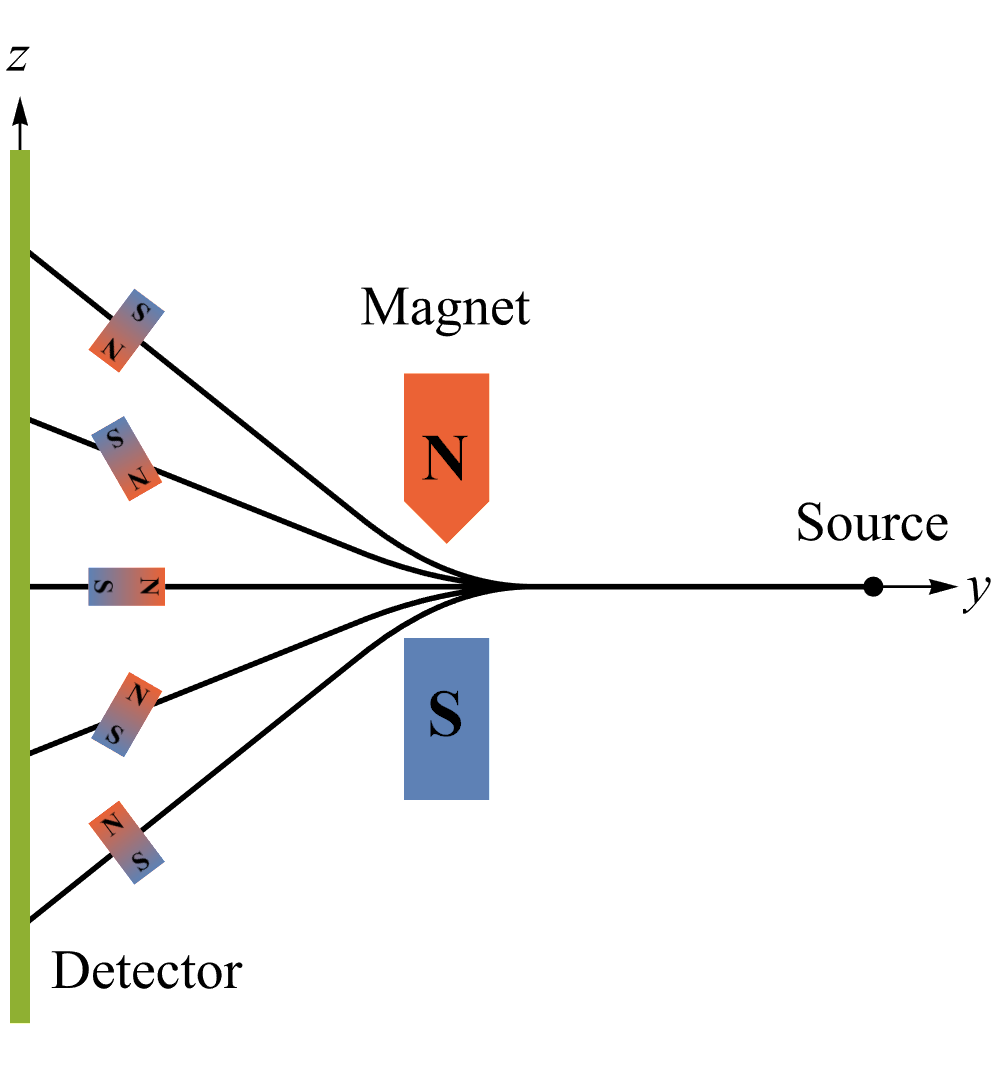}  \caption{The classical model of the Stern-Gerlach experiment. If the atoms enter with random orientations of their `intrinsic' magnetic moments (due to their `intrinsic' angular momenta), the SG magnets should produce all possible deflections.} \label{SGclassical}
\end{center}
\end{figure}

If the classical model of spin was true (Figure \ref{SGclassical}), then Alice or Bob would be measuring a fraction of $h$ when their SG magnets were not aligned. In that case, Alice or Bob's fractional outcome definitely establishes an asymmetry that doesn't allow for the symmetrical temporal flipping of zigzag co-causation. That is, if Bob's(Alice's) outcome is a fraction, then he(she) has to follow Alice's(Bob's) state preparation in a temporally flipped scenario because you can't restore Bob's(Alice's) missing spin angular momentum with a subsequent measurement. Notice that this asymmetry results from violating NPRF. 

\clearpage

Because NPRF holds, the Planck postulate holds and we have the symmetrical quantum outcomes, so Alice and Bob's choices for their SG settings carry equal causal weight. That is, instead of Bob's choice of his SG setting causing the quantum state for Alice to measure or vice-versa, either SG setting can be said to cause the quantum state for the other. Again, ``measurement choices on one side of an Bell experiment \textit{make a difference} to outcomes on the other side, in some cases'' \cite{PriceWhartonColliders2024}. In Section \ref{SectionPrinciple-alternative}, we will show how this symmetry is manifested in Bell state entanglement.


\section{An Acausal Account of Bell State Entanglement}\label{SectionPrinciple-alternative}

Let's revisit the situation depicted in Figure \ref{SGExp2}. The angle between $\hat{z}$ (for $|\psi\rangle = |z+\rangle$) and $\hat{b}$ (for $|b+\rangle$)in Figure \ref{SGExp2} is $\theta$ while the angle between $|z+\rangle$ and $|b+\rangle$ in Hilbert space is $\frac{\theta}{2}$. So, QM says our qubit probabilities are $P(b+|\theta) = \cos^2{\left(\frac{\theta}{2}\right)}$ and $P(b-|\theta) = \sin^2{\left(\frac{\theta}{2}\right)}$ giving an average (expectation value) of 
\begin{center}
    $(+1)\cos^2{\left(\frac{\theta}{2}\right)} + (-1)\sin^2{\left(\frac{\theta}{2}\right)} = \cos{\left(\theta \right)}$.
\end{center}
We can derive these qubit probabilities using NPRF + $h$ and the closeness requirement \cite{dakicBrukner2009} by first demanding
\begin{center}
    $(+1)P(b+|\theta) + (-1)P(b-|\theta) = \cos{\left(\theta \right)}$.
\end{center}
Again, the outcome we expect for the measurement/projection of $|\psi\rangle = |z+\rangle$ along $\hat{b}$, given our classical understanding of angular momentum as a vector, is $\cos{\left(\theta \right)}$ (Figure \ref{Projection}). But, since the measurement at $\hat{b}$ must produce $\pm 1$ just like any other direction in space per NPRF + $h$, we can only \textit{average} this expected projection. And, the closeness requirement \cite{dakicBrukner2009} says that classical properties obtain on average over a large number of their corresponding, coherent quantum states. The quantum-mechanical probabilities for our qubit then follow uniquely from that equation and normalization
\begin{center}
    $P(b+|\theta) + P(b-|\theta) = 1$.
\end{center}
Now let's see how the joint probabilities for a Bell spin state can be derived from NPRF + $h$.

The Bell spin singlet state represents a spin-entangled pair of particles with anti-aligned spins in any direction of space. The Bell states in general represent the quantum conservation of some property; here that property is spin angular momentum. So, when Alice and Bob both happen to make their SG measurements of this state in the same direction of space, they always get opposite outcomes consistent with a total spin angular momentum of zero. In the $z$ basis, it is written
\begin{equation*}
|\psi_-\rangle = \frac{|z+\rangle\otimes|z-\rangle \,- |z-\rangle\otimes|z+\rangle}{\sqrt{2}}
\end{equation*}

\clearpage

\noindent Likewise, the three Bell spin triplet states in the $z$ basis
\begin{center}
$\begin{aligned}
&|\psi_+\rangle = \frac{|z+\rangle\otimes|z-\rangle + |z-\rangle\otimes|z+\rangle}{\sqrt{2}}\\
&|\phi_-\rangle = \frac{|z+\rangle\otimes|z+\rangle \,- |z-\rangle\otimes|z-\rangle}{\sqrt{2}}\\
&|\phi_+\rangle = \frac{|z+\rangle\otimes|z+\rangle + |z-\rangle\otimes|z-\rangle}{\sqrt{2}}\\ \label{BellStates}
\end{aligned}$
\end{center}
represent a pair of spin-entangled particles with aligned spins in any direction of space for their respective symmetry planes. So, when Alice and Bob both measure along the same direction in the relevant plane of symmetry they will always get the same outcomes consistent with a total spin angular momentum of $\pm 2$. Let's look at the correlation functions to clarify that.

If Alice is making her spin measurement $\sigma_1$ in the $\hat{a}$ direction and Bob is making his spin measurement $\sigma_2$ in the $\hat{b}$ direction (Figure \ref{EPRBmeasure}), we have
\begin{equation}
\begin{aligned}
    &\sigma_1 = \hat{a}\cdot\vec{\sigma}=a_x\sigma_x + a_y\sigma_y + a_z\sigma_z \\
    &\sigma_2 = \hat{b}\cdot\vec{\sigma}=b_x\sigma_x + b_y\sigma_y + b_z\sigma_z \\ \label{sigmas}
\end{aligned}
\end{equation}
where $(\sigma_x,\sigma_y,\sigma_z)$ are the Pauli spin operators. The correlation functions are then given by
\begin{equation}
\begin{aligned}
&\langle\psi_-|\sigma_1\sigma_2|\psi_-\rangle = &-a_xb_x - a_yb_y - a_zb_z\\
&\langle\psi_+|\sigma_1\sigma_2|\psi_+\rangle = &a_xb_x + a_yb_y - a_zb_z\\
&\langle\phi_-|\sigma_1\sigma_2|\phi_-\rangle = &-a_xb_x + a_yb_y + a_zb_z\\
&\langle\phi_+|\sigma_1\sigma_2|\phi_+\rangle = &a_xb_x - a_yb_y + a_zb_z\\ \label{gencorrelations}
\end{aligned}
\end{equation}

\noindent You can see that the correlation function for the singlet state is $-\cos{\left(\theta \right)}$ while the correlation function for a triplet state is $\cos{\left(\theta \right)}$ in a particular plane, e.g., that symmetry plane for $|\psi_+\rangle$ is the $xy$ plane. Let's look at a triplet state in its symmetry plane.

\begin{figure}
\begin{center}
\includegraphics [height = 60mm]{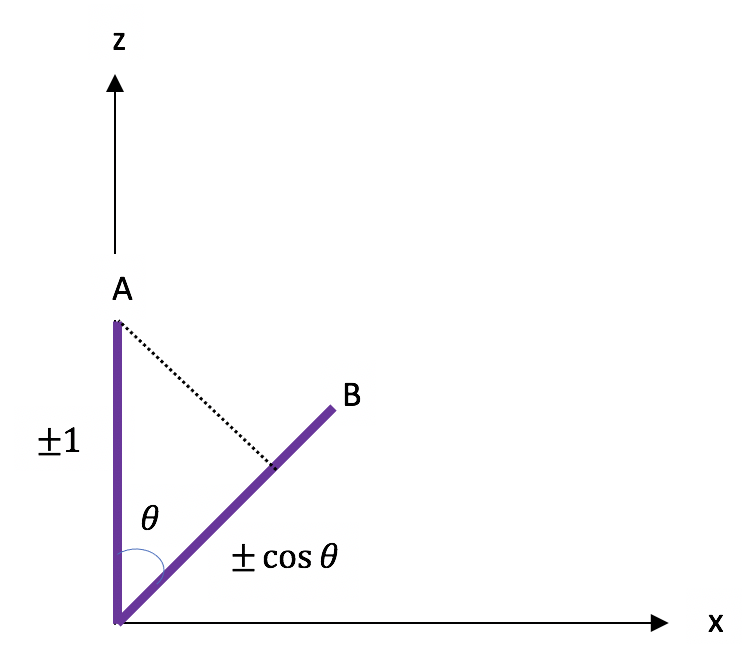}  \caption{Per Alice, Bob should be measuring $\pm \cos{\left(\theta \right)}$ when she measures $\pm1$, respectively.} \label{Alice-View}
\end{center}
\end{figure}

\begin{figure}
\begin{center}
\includegraphics [width = \textwidth]{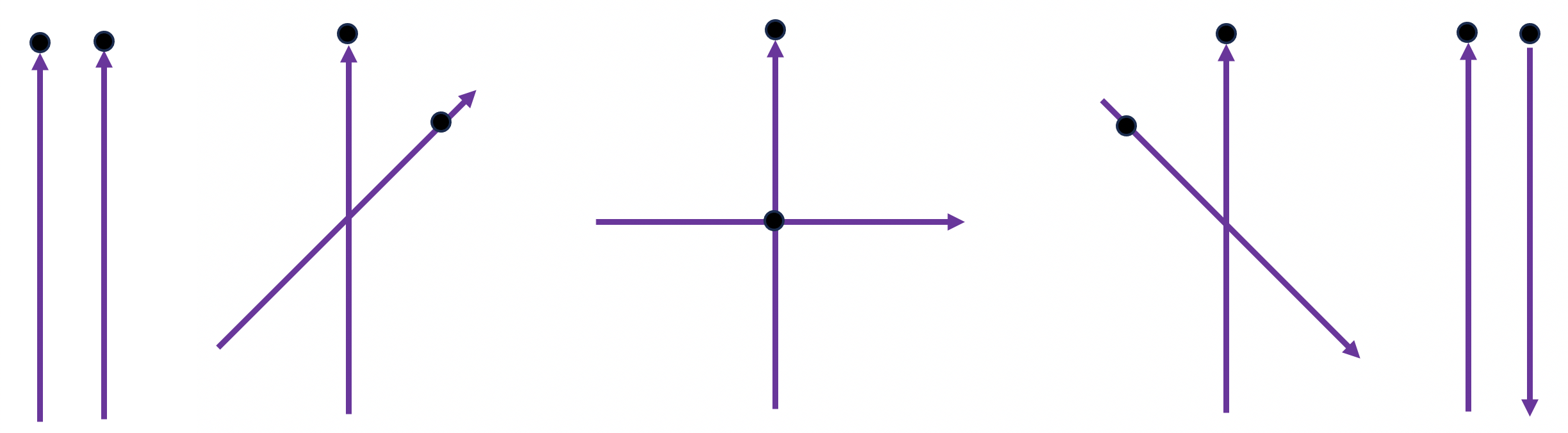}  \caption{Average View for the Triplet State. Reading from left to right, as Bob rotates his SG magnets (rotating purple arrow) relative to Alice's SG magnets (purple arrow always vertically oriented) for her $+1$ outcome (black dot at tip of her arrow), the average value of his outcome (black dot along his arrow) varies from $+1$ (totally up, arrow tip) to $0$ to $-1$ (totally down, arrow bottom). This obtains per conservation of spin angular momentum on average in accord with NPRF. Bob can say exactly the same about Alice's outcomes as she rotates her SG magnets relative to his SG magnets for his $+1$ outcome.} \label{AvgViewTriplet}
\end{center}
\end{figure}

\clearpage

Suppose Alice obtains $+1$ at $\hat{a}$ and Bob measures at $\hat{b} \ne \hat{a}$ ($\theta \ne 0$). We have the exact same situation here between $\hat{a}$ and $\hat{b}$ for two particles that we had above between $\hat{z}$ and $\hat{b}$ for one particle. Using the same reasoning here, Alice says Bob's measurement outcome should be $\cos{\left(\theta \right)}$, since obviously he would have also gotten $+1$ for his particle if he had measured at $\hat{b} = \hat{a}$, as required to conserve spin angular momentum (Figure \ref{Alice-View}). The problem is, that would mean Alice alone measures $h$ while Bob measures some fraction of $h$, which means Alice occupies a preferred reference frame. Since Bob must also always measure $h$ per NPRF, Bob's $\pm 1$ outcomes can only \textit{average} to $\cos{\left(\theta \right)}$ at best (Figure \ref{AvgViewTriplet}). That means from Alice's perspective, Bob's measurement outcomes only satisfy conservation of spin angular momentum \textit{on average} when Bob is measuring the spin\index{quantum spin} of his particle in a different inertial reference frame. 

\begin{figure}
\begin{center}
\includegraphics [height = 60mm]{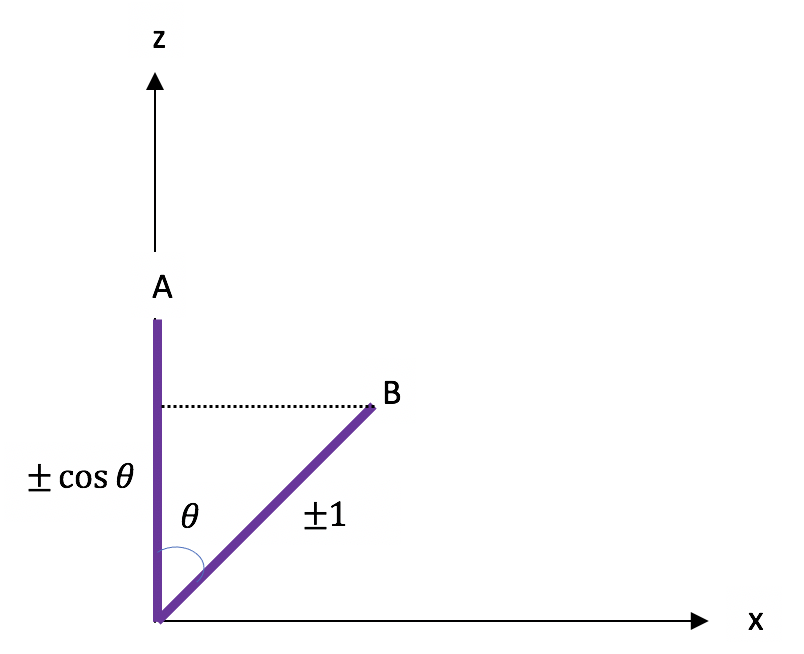}  \caption{Per Bob, Alice should be measuring $\pm \cos{\left(\theta \right)}$ when he measures $\pm1$, respectively.} \label{Bob-View}
\end{center}
\end{figure}

We can write this `average-only' conservation\index{`average-only' conservation} for Alice's $+1$ outcomes as 
\begin{center}
$2P(++)(+1) + 2P(+-)(-1) = \cos{\left(\theta \right)}$.
\end{center}
Likewise, for Alice's $-1$ outcomes `average-only' conservation is written
\begin{center}
$2P(-+)(+1) + 2P(--)(-1) = -\cos{\left(\theta \right)}$.
\end{center}
This `average-only' conservation plus normalization per NPRF
\begin{align*}
P(++) + P(+-) & = \frac 12 \\
P(-+) + P(--) & = \frac 12
\end{align*}
gives precisely the QM joint probabilities $P(++) = P(--) = \frac{1}{2}\cos^2{\left(\frac{\theta}{2}\right)}$ and $P(+-) = P(-+) = \frac{1}{2}\sin^2{\left(\frac{\theta}{2}\right)}$. It is a simple matter to repeat this derivation for the single state and obtain $P(++) = P(--) = \frac{1}{2}\sin^2{\left(\frac{\theta}{2}\right)}$ and $P(+-) = P(-+) = \frac{1}{2}\cos^2{\left(\frac{\theta}{2}\right)}$. Of course, Bob can make the exact same argument about Alice's outcomes (Figure \ref{Bob-View}) and derive the joint probabilities from his perspective. 

\begin{figure}
\begin{center}
\includegraphics [width = \textwidth]{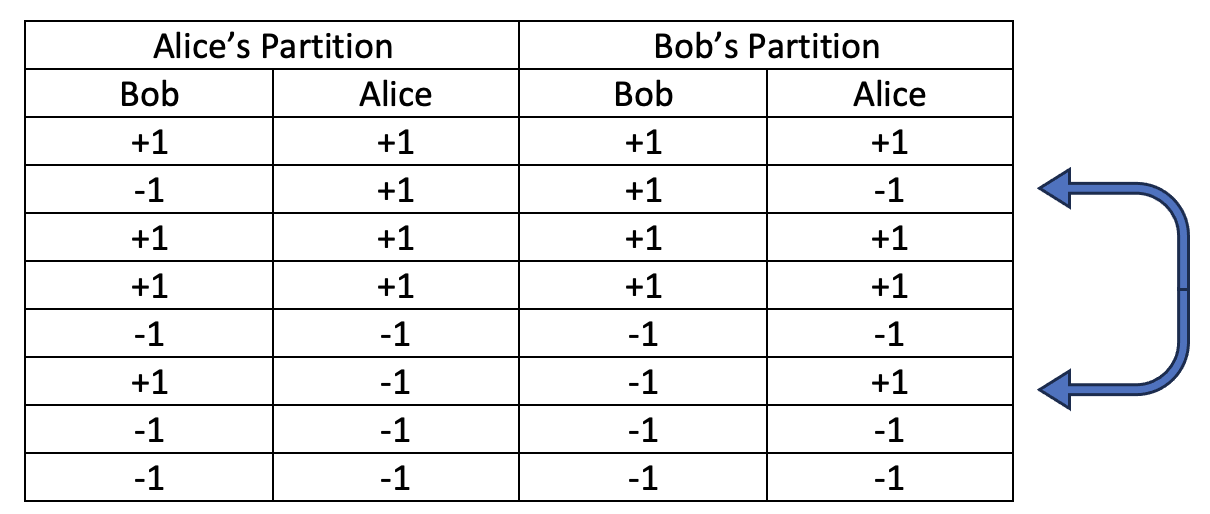}  \caption{Example collection of eight data pairs when Alice and Bob's measurement settings in the symmetry plane for a Bell spin triplet state differ by $60^{\circ}$. Alice partitions the data according to her $\pm 1$ results to show that Bob's measurement outcomes only \textit{average} the required $\pm\frac{1}{2}$ for conservation of spin angular momentum. But, when Bob partitions the data according to his $\pm 1$ results (by switching rows 2 and 6 shown with blue arrows) he can show it is \textit{Alice's} measurement outcomes that only average the required $\pm\frac{1}{2}$ for conservation of spin angular momentum.} \label{AliceBobData}
\end{center}
\end{figure}

In this view, the mystery of Bell state entanglement resides in the symmetric, `average-only' conservation that results from the observer-independence of $h$ between inertial reference frames related by spatial rotations (Figure \ref{AliceBobData}). Accordingly, the mystery is easily solved by justifying the empirically discovered fact responsible for the mystery with the relativity principle, NPRF + $h$. 

\section{Analogy with Special Relativity}\label{SectionSR}

This is totally analogous to the mystery of length contraction resulting from the observer-independence of $c$ between inertial reference frames related by boosts. There, Alice's measurements show clearly that Bob's meter sticks are shorter than hers while Bob's measurements show clearly that Alice's meter sticks are shorter than his. 

\begin{figure}
\begin{center}
\includegraphics [width=\textwidth]{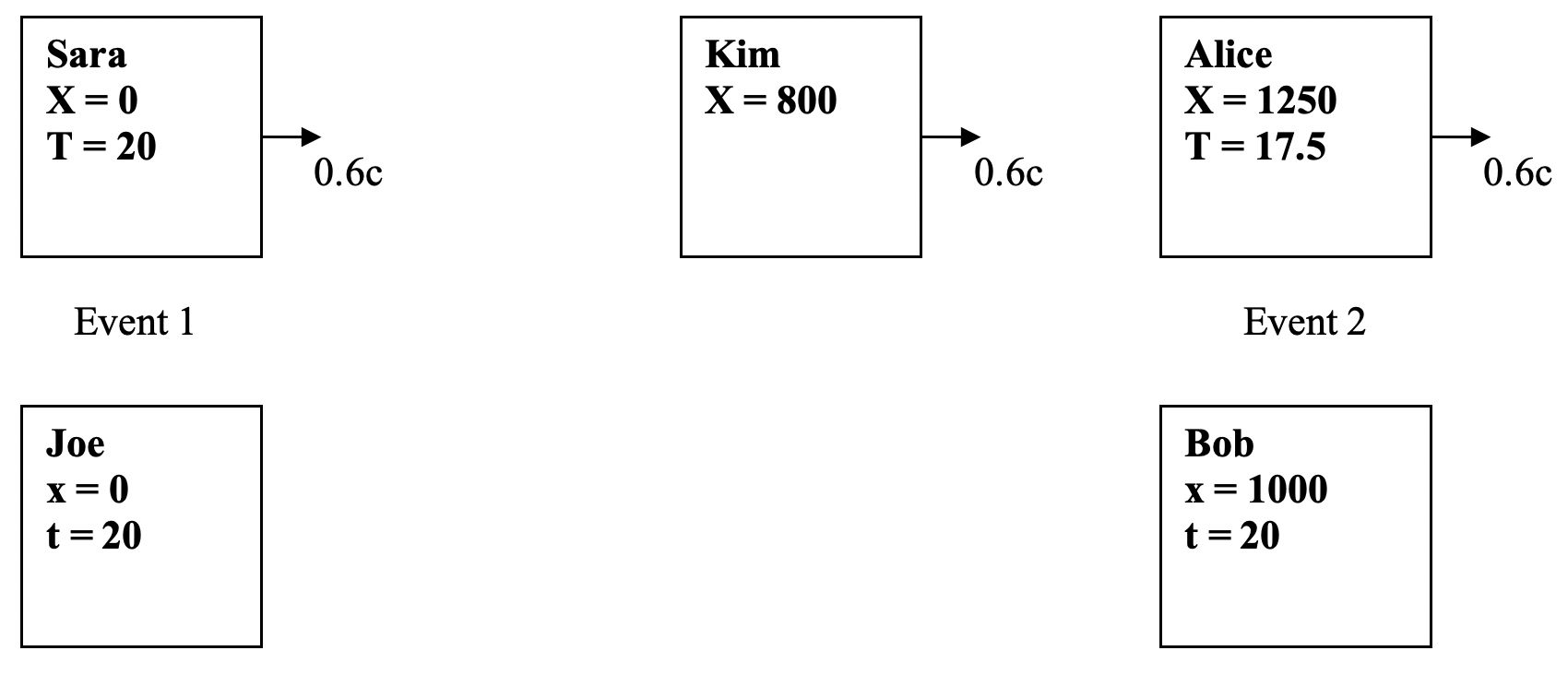}  \caption{The girls are the same age in their reference frame and the boys are the same age in their reference frame. We exaggerate the time differences for effect. The temporal origin corresponds to being 20 years old and -0.0025 s corresponds to being 2.5 years younger than 20 years old, i.e., 17.5 years old, and so on. The distances are in km. This figure shows Events 1 and 2 occurring simultaneously from the boys' perspective, which leads them to conclude that the girls are not the same age and their meter sticks are clearly shorter.} \label{RoS1}
\end{center}
\end{figure}

\begin{figure}
\begin{center}
\includegraphics [width=\textwidth]{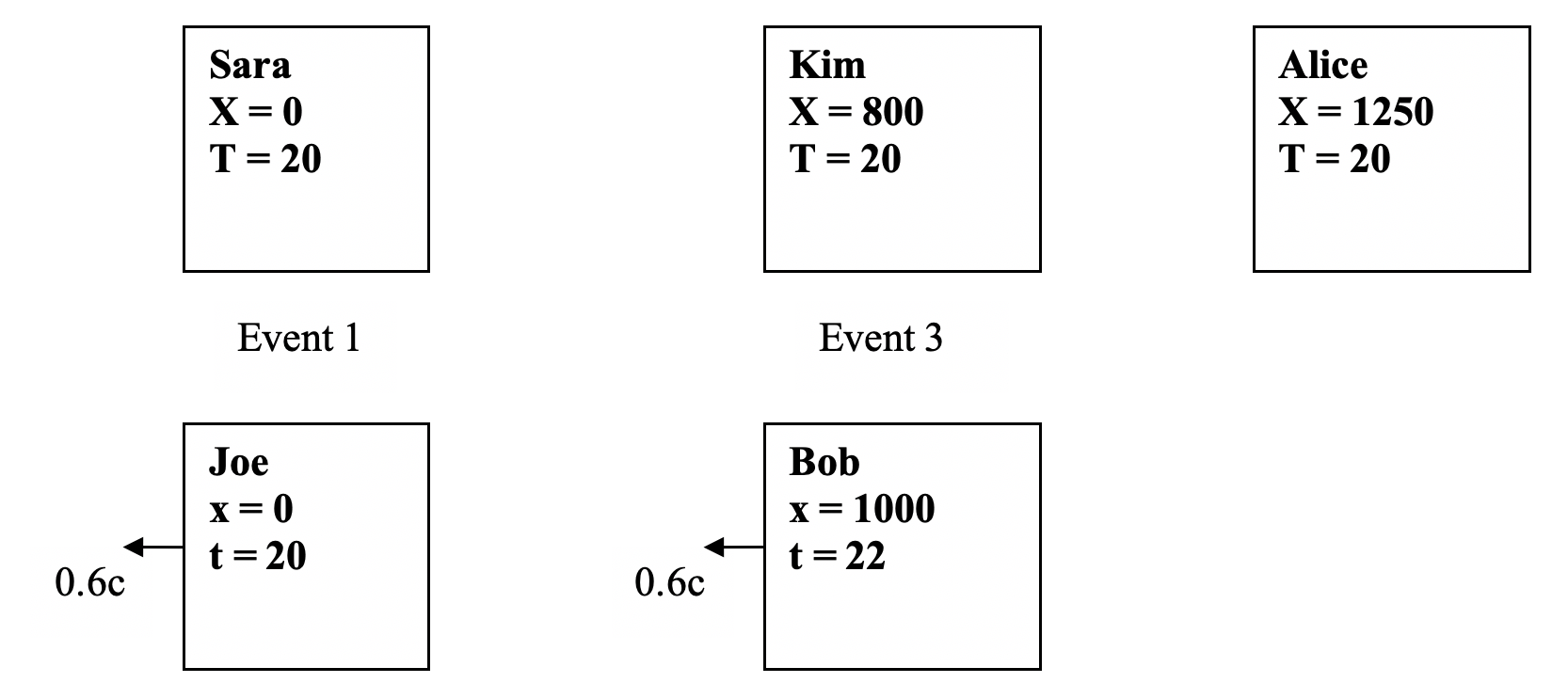}  \caption{The girls say Events 1 and 3 are simultaneous, so the boys are not the same age and their meter sticks are clearly shorter.} \label{RoS3}
\end{center}
\end{figure}

For example, consider the following three events adapted from the 2004 version by DeWitt \cite{dewitt} (who adapted his version from Mermin) using exaggerated time differences: 
\begin{itemize}
\item Event 1: 20 year-old Joe and 20 year-old Sara meet.
\item Event 2: 20 year-old Bob and 17.5 year-old Alice meet.
\item Event 3: 22 year-old Bob and 20 year-old Kim meet.
\end{itemize}
The girls and the boys agree on the facts contained in these three events. Further, Joe and Bob see the girls moving in the positive $x$ direction (Figure \ref{RoS1}), so the girls see the boys moving in the negative $X$ direction at the same speed (Figure \ref{RoS3}). Additionally, the boys are the same age in their reference frame and the girls are the same age in their reference frame. This establishes simultaneity for each set, i.e., events are simultaneous (coexist) for the boys if the events occur when the boys are the same age, e.g., Events 1 and 2 above. Likewise for the girls, e.g., Events 1 and 3 above. This is known as the \textit{relativity of simultaneity} and motivates the block universe interpretation of Minkowski spacetime.

\clearpage

To appreciate this, we note that Einstein specifically defined time using simultaneity \cite{SR1905}:
\begin{quote}
We have to take into account that all our judgments in which time plays a part are always judgments of \textit{simultaneous events}. [italics in original]
\end{quote}
\noindent where his notion of simultaneity was that of the synchronicity of \textit{stationary} clocks, which was established by exchanging light signals \cite{SR1905}:
\begin{quote}
Thus with the help of certain imaginary physical experiments [associated with the exchange of light signals] we have settled what is to be understood by synchronous stationary clocks located at different places, and have evidently obtained a definition of ``simultaneous,'' or ``synchronous,'' and of ``time.'' The ``time'' of an event is that which is given simultaneously with the event by a stationary clock located at the place of the event, this clock being synchronous, and indeed synchronous for all time determinations, with a specified stationary clock. ... \\

\noindent It is essential to have time defined by means of stationary clocks in the stationary system, and the time now defined being appropriate to the stationary system we call it ``the time of the stationary system.''
\end{quote}
\noindent So, the boys' clocks (mechanical and biological) are stationary and synchronized with respect to each other and establish ``the time of [their] stationary system'' while the same is true for the girls' clocks (mechanical and biological). Accordingly, the boys are the same age in their stationary system and say they coexist when they are the same age while the girls are the same age in their stationary system and would make the same claim about their coexistence. 

\begin{figure}
\begin{center}
\includegraphics [height = 120mm]{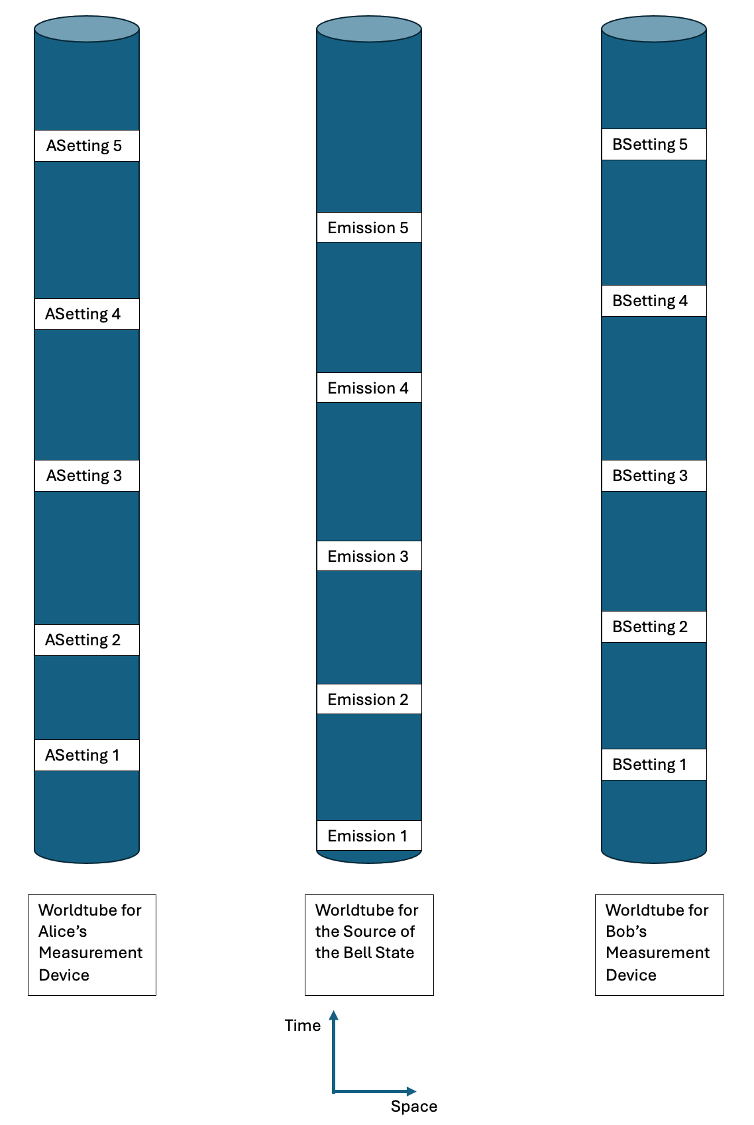}  \caption{Alice and Bob choose their SG measurement settings randomly and independently for each trial/emission event of the experiment and the Bell state source is faithfully reproduced in each trial.} \label{AGC1}
\end{center}
\end{figure}

\begin{figure}
\begin{center}
\includegraphics [height = 120mm]{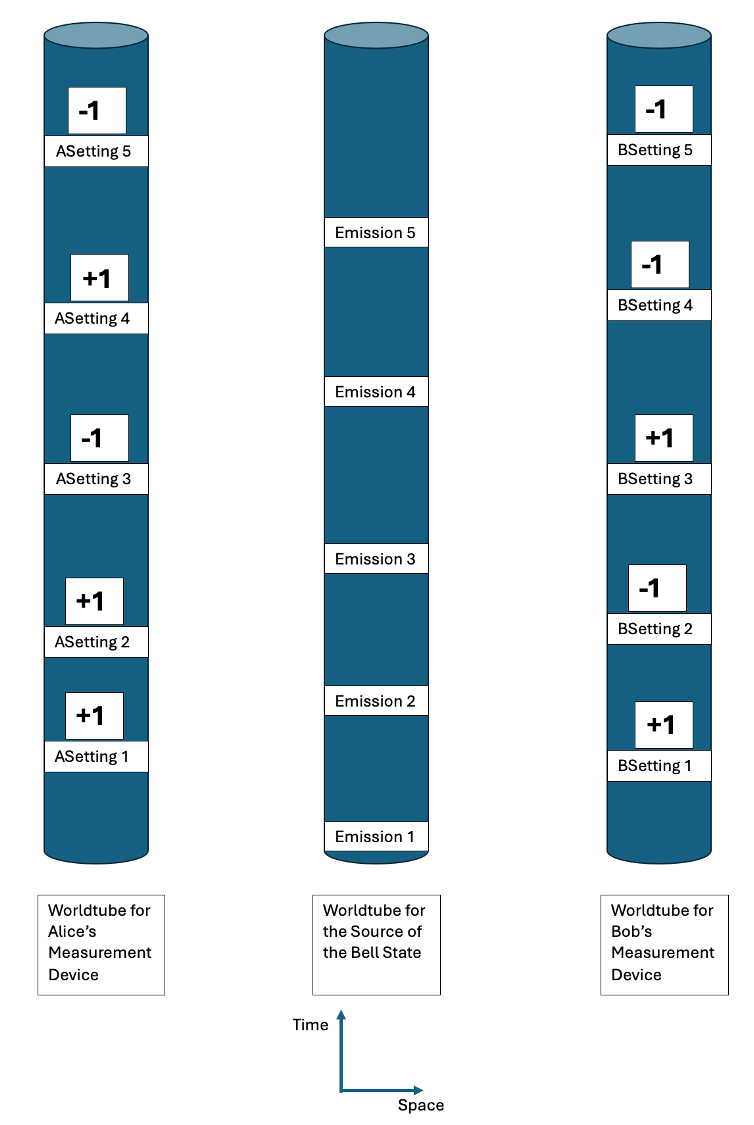}  \caption{The joint probabilities per our acausal global constraint NPRF + $h$ provide an all-at-once distribution of outcomes throughout spacetime for Alice and Bob's SG measurement choices over the Bell state source in question.} \label{AGC2}
\end{center}
\end{figure}

This motivated Geroch to write \cite[p. 20--21]{geroch}:
\begin{quote}
There is no dynamics within space-time itself: nothing ever moves therein; nothing happens; nothing changes. In particular, one does not think of particles as moving through space-time, or as following along their world-lines. Rather, particles are just in space-time, once and for all, and the world-line represents, all at once, the complete life history of the particle.
\end{quote}
Consequently, we introduced the term ``all at once'' explanation in \cite{stuckeyFoP} to describe the 4D global distribution of quantum outcomes among the 4D configuration of worldtubes for the experimental equipment (beam splitters, mirrors, sources, SG magnets, detectors, etc.). Accordingly, NPRF + $h$ is an adynamical global constraint on the distribution of quanta in the worldtubes of the experimental equipment configured per the adynamical global constraint NPRF + $c$ (Figures \ref{AGC1} and \ref{AGC2}). 

\clearpage

Thus, there is no need to invoke causal mechanisms to explain the paradoxical kinematic facts of length contraction and `average-only' conservation. These kinematic facts follow necessarily from the adynamical global constraints NPRF + $c$ and NPRF + $h$, respectively. Now let's compare this principle account of Bell state entanglement with CCC per Price and Wharton.

\section{Co-Causation or No Causation?}\label{SectionCo-causation}

If we think of Figure \ref{ZigZag1A} as Alice's subjective spacetime model of reality and Figure \ref{ZigZag1} as Bob's subjective spacetime model of reality according to causal perspectivalism, then what happens to the zigzag in the objective spacetime model of reality? The combined effect of both zigzags amounts to what Wharton and Liu called a ``causally-neutral account,'' so the zigzags are functionally superfluous in the objective spacetime model of reality. By simply omitting the zigzags in the objective spacetime model of reality, we could avoid causal talk altogether and conclude that NPRF + $h$ provides an acausal global constraint explanation of Bell state entanglement that does not violate locality, statistical independence, intersubjective agreement, or unique experimental outcomes. It is only when we attempt to further underwrite NPRF + $h$ via zigzag paths in the objective spacetime model of reality that we introduce constructive explanation and violate statistical independence. It is the desire to keep constructive explanation, causal explanation, etc., in our objective spacetime model of reality at all costs that gets us into trouble when explaining Bell-inequality-violating correlations. 

So, we have violated statistical independence by trying to explain the all-at-once retrocausality of the objective spacetime model of reality as the result of two timelike causal zigzag paths from the two subjective spacetime models of reality according to causal perspectivalism. Notice the analogy with length contraction above. 

There, from Alice's perspective (her reference frame), Bob's meter sticks are short while from Bob's perspective (his reference frame), Alice's meter sticks are short. Here, Alice can claim it is her choice of measurement setting that is causally influencing Bob's measurement outcome (Figure \ref{ZigZag1A}), while Bob can say the same thing about his choice of measurement setting causally influencing Alice's measurement outcome (Figure \ref{ZigZag1}). Is it better to think of this as quantum co-causation? Or, should we rather think of it as no causation? Let's look again at SR to help us answer that question. 

If Alice's perspective (reference frame) is preferred, then there must be some causal mechanism literally shortening Bob's meter sticks. As a result, Bob measures the same value $c$ for the speed of a light beam that Alice does, even though they are moving relative to each other. Conversely, if Bob's perspective (reference frame) is preferred, the causal mechanism is actually shortening Alice's meter sticks, again resulting in her measuring the same value $c$ for the speed of a light beam that Bob does. What is the consensus attitude about this in physics? 

There is no causal mechanism shortening meter sticks in any reference frame. Alice's measurements indicate that Bob's meter sticks are short while Bob's measurements indicate the same thing about Alice's meter sticks, precisely because neither perspective is preferred (NPRF). The relativity principle means Alice and Bob must both measure the same value $c$ for the speed of a light beam and that leads to the relativity of simultaneity, which explains Alice and Bob's measurement outcomes without having to invoke causal mechanisms at all. Length contraction is a kinematic fact, not a dynamical effect. 

Extrapolating the principle approach from SR to the quantum situation here suggests we again drop the notion of different causal influences invoked from different perspectives, i.e., different causal zigzag patterns invoked from their different reference frames. If there is some causal mechanism responsible for their disagreement, it is responsible for Bob measuring the same value $h$ for Planck's constant that Alice does, even though their measurement devices are oriented in different directions. 

So, following the analogous situation in SR, we might simply adopt the relativity principle to justify the empirical fact that Alice and Bob both measure the same value $h$ for Planck's constant. That leads to `average-only' conservation, which explains Alice and Bob's measurement outcomes without having to invoke causal mechanisms at all. `Average-only' conservation is a kinematic fact, not a dynamical effect.  Consequently, this NPRF + $h$ principle explanation does not violate locality, statistical independence, intersubjective agreement, or unique experimental outcomes.


\section{Alternative View of CCC}\label{SectionConcl}

With all that being said, this need not be the last word concerning CCC. If the CCC advocate accepts the completeness of QM, the zigzags in the subjective spacetime models are no longer associated with (hidden) causal mechanisms. The zigzags are simply statements of perspectival causation, i.e., Alice and Bob's settings independently ``make a difference'' for the Bell-inequality-violating correlations when the source is a Bell state. Since the Bell states are complete, and Alice and Bob choose their measurement settings randomly and independently for the Bell state being measured, there is no violation of statistical independence. And, since the zigzags follow timelike paths, there is no violation of locality for this perspectival causation. 

We simply need to identify the mechanism motivating us to invoke the co-causation of CCC. That is, when the source is not a Bell state, but a simple source of classical particles with counterfactually definite properties, no one would invoke the subjective co-causation of CCC; because in that case, we don't violate Bell's inequality and the causal explanation of the correlations follows the blue arrows in Figure \ref{Vshape} (for non-spin, classical particles) trivially. So, contrary to Price and Wharton, we see that CCC is not a mechanism per se, it's a causal account of some mechanism per causal perspectivalism. Why are we motivated to invoke CCC? The answer is simple, the source is a Bell state, that's the constraint for our collider that motivates us to invoke subjective co-causation per CCC. In this case, CCC is understood as the subjective, causal counterpart to the objective, acausal NPRF + $h$ and there is no violation of locality or statistical independence.

If we want to solve the mystery of quantum entanglement in accord with Reichenbach's Principle, then we must violate locality, statistical independence, intersubjective agreement, and/or unique experimental outcomes. This leads to a problem articulated by Van Camp \cite{vancamp2011}, ``Constructive interpretations are attempted, but they are not unequivocally constructive in any traditional sense.'' He concludes \cite{vancamp2011}:
\begin{quote}
    The interpretive work that must be done is less in coming up with a constructive theory and thereby explaining puzzling quantum phenomena, but more in explaining why the interpretation counts as explanatory at all given that it must sacrifice some key aspect of the traditional understanding of causal-mechanical explanation.
\end{quote}
If statistical mechanics is the paradigm example of constructive explanation, then in light of Bell's theorem and the experimental violation of Bell's inequality, it is hard to imagine any constructive account of quantum mechanics gaining consensus support. And, if SR (following from NPRF + $c$) is the paradigm example of principle explanation without a consensus constructive counterpart, it is easy to imagine QM (following from NPRF + $h$) also remaining without a consensus constructive counterpart.

There is already a definite trend towards atemporal aka all-at-once \cite{stuckeyFoP} aka acausal global constraint explanation for Bell-inequality-violating correlations. Adlam and Rovelli write \cite{AdlamRovelli}:
\begin{quote}
Moreover, ... we do not need to think of the set of events as being generated in some particular temporal order. In fact, we can say something even stronger: if we want to maintain relativistic covariance then we cannot think of the set of events as being generated in some particular temporal order. This point has been noted in the context of other ontologies consisting of pointlike events - for example, Esfeld and Gisin note that the Bell flash ontology\index{Bell flash ontology} is relativistically covariant only if ``one limits oneself to considering whole possible histories or distributions of flashes in spacetime, and one renounces an account of the temporal development of the actual distribution of the flashes in space-time.'' [26] Thus it seems that [relational quantum mechanics]\index{relational quantum mechanics} is most compatible with a metaphysical picture in which the laws of nature apply atemporally to the whole of history, fixing the entire distribution of quantum events all at once.
\end{quote}

\clearpage

\noindent They note that this kind of explanation does not ``involve hidden influences or preferred reference frames, and thus there is no particular reason to try to avoid this sort of nonlocality\footnote{Here ``nonlocality'' is referring to relations between events distributed over large regions of spacetime as opposed to small (local) regions of spacetime. To avoid confusion with nonlocality per superluminal causal connections, the better term for this type of ``nonlocality'' is probably ``non-separability'' \cite{SalomBell2023}.}.'' Concerning the laws of physics more generally, Chen and Goldstein write \cite{ChenGoldstein2021}:
\begin{quote}
It is sometimes assumed that the governing view of laws requires a fundamental direction of time: to govern, laws must be \textit{dynamical} laws that \textit{produce} later states of the world from earlier ones, in accord with the direction of time that makes a fundamental distinction between past and future. ... On our view, fundamental laws govern by constraining the physical possibilities of the entire spacetime and its contents. They need not exclusively be dynamical laws, and their governance does not presuppose a fundamental direction of time. For example, they can take the form of global constraints or boundary-condition constraints for spacetime as a whole;
\end{quote}
This is precisely how one should view the principle constraints NPRF + $h$ and NPRF + $c$. However, this ignores the dynamical nature of our subjective experience, which provides the subjective spacetime models of reality whence the objective spacetime model. Therefore, it seems reasonable to augment NPRF + $h$ and NPRF + $c$ with a complementary, subjectively causal counterpart. CCC properly understood can provide that counterpart.

\bibliographystyle{siam}
\bibliography{biblio} 
\end{document}